\documentclass[aps,prd,11pt,notitlepage,longbibliography,nofootinbib,tightenlines,preprintnumbers]{revtex4-1}

\pdfoutput=1
\usepackage{amsmath,amssymb,amsfonts}
\usepackage{graphicx}
\usepackage{color}
\usepackage{tikz}
\usepackage{dsfont}
\usepackage[pdftex]{hyperref}
\hypersetup{colorlinks=true, linkcolor=darkred, citecolor=blue, linktoc=page}
\definecolor{darkred}{rgb}{0.8,0.1,0.1}

\usepackage{diagbox}

\usepackage{subfigure}

\def\RR{{\mathds{R}}}

\def\ZZ{{\mathds{Z}}}

\DeclareMathOperator{\vol}{vol}
\DeclareMathOperator{\Vol}{Vol}

\makeatletter\def\l@subsubsection#1#2{}%
\makeatother

\newcommand{\nocontentsline}[3]{}
\newcommand{\tocless}[2]{\bgroup\let\addcontentsline=\nocontentsline#1{#2}\egroup}

\begin{document}

\title{Information transfer with a twist}

\author{Christoph F.~Uhlemann} 
\email{uhlemann@umich.edu}

\affiliation{Leinweber Center for Theoretical Physics, Department of Physics	\\
	University of Michigan, Ann Arbor, MI 48109-1040, USA}

\preprint{LCTP-21-33}

\begin{abstract}
Holographic duals for CFTs compactified on a Riemann surface $\Sigma$ with a twist are cast in the language of wedge holography. $\Sigma$ starts as part of the field theory geometry in the UV and becomes part of the internal space in the IR. This allows to associate entanglement entropies with splits of the internal space in the IR geometry. Decomposing the internal space in the IR and geometrizing the corresponding subsystems separately leads to two interacting gravitational systems, similar to the intermediate holographic description in braneworld models. For $\Sigma=T^2$ the setups are used to model information transfer from a black hole to a gravitating bath. This leads to Page curves with a phase structure which precisely mirrors that in braneworld models. The transition from geometric to non-geometric entropies is also discussed for $\Sigma=S^2$ as a model for more general internal spaces in AdS/CFT.
\end{abstract}

\maketitle
\tableofcontents
\parskip 1mm

\section{Introduction and summary}

Recent progress in the understanding of black holes includes the accounting for the microstates of supersymmetric black holes through indices in holographically dual QFTs on the one hand, and an improved understanding of the black hole radiation entropy on the other.
A canonical example in the former context are magnetically charged black holes in AdS, whose microstates can be accounted for by the topologically twisted index of the dual QFTs \cite{Benini:2015noa,Benini:2015eyy}. 
For the entropy of Hawking radiation recent work \cite{Penington:2019npb,Almheiri:2019psf,Almheiri:2019hni,Almheiri:2019yqk,Rozali:2019day,Almheiri:2019psy,Almheiri:2019qdq,Penington:2019kki} has shown that unitary Page curves can be obtained from semi-classical gravity, building on the notion of quantum extremal surfaces \cite{Engelhardt:2014gca,Faulkner:2013ana}.
Reviews can be found in \cite{Zaffaroni:2019dhb} and \cite{Almheiri:2020cfm,Raju:2020smc}.
Here we will connect these two lines of research.

The Page curve studies are largely based on doubly-holographic Karch/Randall braneworld models \cite{Karch:2000gx,Karch:2000ct} (see also \cite{Takayanagi:2011zk,Fujita:2011fp}), in which quantum extremal surfaces can be studied by means of classical Ryu/Takayanagi surfaces.
Page curves have been obtained in two ways: (i) one can couple a gravitating black hole system to a non-gravitating bath and study information transfer to the non-gravitating system; this makes the graviton massive \cite{Geng:2020qvw} (ii) one can couple a black hole to a gravitating bath, with black hole and bath systems defined by a non-geometric split of the degrees of freedom; this was realized in braneworld models in \cite{Geng:2020fxl}.
Both scenarios were realized for four-dimensional black holes in UV-complete string theory settings in \cite{Uhlemann:2021nhu}, which confirmed the emergence of Page curves. 
In scenario (ii) there can be a massless graviton in the combined system, but each subsector only has access to massive modes \cite{Geng:2021hlu}.
That way Page curves in both setups are compatible with the arguments of \cite{Laddha:2020kvp,Raju:2021lwh} to the effect that in theories with dynamical long-range gravity all information is available outside the black hole at all times.

Here we study holographic duals of CFTs compactified on a Riemann surface $\Sigma$.
This leads to RG flows from a $(d+2)$-dimensional CFT$_{d+2}$ on $R^{d}\times \Sigma$ in the UV to a $d$-dimensional CFT$_d$ in the IR. 
Supersymmetry can be preserved through a partial topological twist.
We focus on the case $\Sigma=T^2$ to realize black holes coupled to a gravitating bath, and study information transfer from the black hole to the bath.
The models are in category (ii); they are embedded in string theory and share qualitative features with the string theory  setups employed for category (ii) in \cite{Uhlemann:2021nhu}. 
We also discuss the transition from geometric entanglement entropy (EE) in the UV to non-geometric entropies in the IR for $\Sigma=S^2$ as a model for more general internal spaces in AdS/CFT.

In string theory terms  the brane configurations which engineer the CFT$_{d+2}$ are wrapped on the Riemann surface $\Sigma$, along the lines of \cite{Maldacena:2000mw}.
The associated 10d or 11d supergravity solutions interpolate between an $AdS_{d+3}$ solution in the UV and an $AdS_{d+1}$ solution in the IR.
However, the features relevant here can for most examples be described conveniently within consistent truncations to $(d+3)$-dimensional gauged supergravities, where the flows take the simple form
\begin{align}
AdS_{d+3}\qquad \longrightarrow\qquad AdS_{d+1}\times \Sigma~.
\end{align}
As concrete examples we discuss compactifications of $\mathcal N=4$ SYM, corresponding to D3-branes wrapped on $\Sigma$ and $d=2$,
and compactifications of the 6d $\mathcal N=(2,0)$ theory, corresponding to M5-branes wrapped on $\Sigma$ and $d=4$.
Microscopic realizations for other $d$ will be discussed briefly.

The setups can be cast in the language of the wedge holography of \cite{Akal:2020wfl}.
Wedge holography starts with AdS$_{d+2}$ cut off by two end-of-the-world (ETW) branes (fig.~\ref{fig:wedge}), as holographic dual for a CFT$_{d+1}$ on an interval.
In the IR limit the interval at the conformal boundary of AdS$_{d+2}$ is reduced to a point.
The CFT$_{d+1}$ reduces to a CFT$_d$ and the interval becomes the internal space in the holographic dual. 
Compactifications on a Riemann surface $\Sigma$ are analogous:
Starting point is a CFT$_{d+2}$ on $\RR^{d}\times \Sigma$, whose dual is an asymptotically-$AdS_{d+3}$ space with conformal boundary $\RR^{d}\times \Sigma$. In the IR the CFT reduces its dimension by two.
In the dual geometry $\Sigma$ becomes the internal space (fig.~\ref{fig:torus-wedge}) and the product of $\Sigma$ and the $AdS_{d+1}$ radial coordinate is a 3d analog of the 2d wedge.
The transition from  AdS$_{d+3}$/CFT$_{d+2}$ to AdS$_{d+1}$/CFT$_d$ may be seen as codimension-3 holography, similar to the interpretation of wedge holography as codimension-2 holography.

The RG flow perspective provides a way to associate EE with splits of the internal space in the IR geometry:
In the UV $\Sigma$ is part of the field theory geometry. Splitting it amounts to a geometric split of the CFT Hilbert space and the associated EE can be computed using the Ryu/Takayanagi prescription. Upon flowing to the IR, $\Sigma$ becomes the internal space. 
The Ryu/Takayanagi surface becomes a surface splitting the internal space in the IR geometry, and the geometric split turns into a non-geometric split of the CFT Hilbert space according to where degrees of freedom are represented on the internal space.
Analogous arguments were used in braneworld models to define subsectors associated with the left and right ETW branes  \cite{Geng:2020fxl} (see also \cite{Geng:2021iyq}). 
In the string theory setups discussed in \cite{Uhlemann:2021nhu}, in which a full internal space supersedes the wedge region, the left/right split becomes a split in the internal space as well.

Decomposing the IR CFT$_d$ degrees of freedom according to their representation on the internal space defines two subsectors.
We can geometrize the two subsectors separately, leading to two asymptotically-$AdS_{d+1}$ holographic duals.
The two gravitating $AdS_{d+1}$ systems are linked by interactions at the boundary that arise from CFT interactions across the cuts that split the internal space.
This is analogous to the `intermediate' holographic description in braneworld models, which can be understood similarly: One starts with a CFT$_{d+1}$ on an interval which is at both ends coupled to $d$-dimensional boundary degrees of freedom, represented by the ETW branes.
Geometrizing only the $d$-dimensional boundary degrees of freedom leads to two asymptotically $AdS_{d+1}$ duals which are coupled through the CFT$_{d+1}$ `bridge'.
In the twisted compactifications the size of $\Sigma$ controls the total number of degrees of freedom, and the decomposition controls how they are split between the two subsystems; this corresponds to the brane angles in the braneworld models. 

We start with the setups with $\Sigma=T^2$ and use them to model information transfer from a black hole to a gravitating bath.
In the IR $AdS_{d+1}\times T^2$ solution, black holes can be realized by replacing the $AdS_{d+1}$ factor by a planar $AdS_{d+1}$ black hole.
From the UV perspective this introduces a temperature which is small compared to the compactification scale, so that the CFT$_{d+2}$  first flows to a CFT$_d$ and then to a thermal state.
We can divide the system into a black hole and a bath system by decomposing one $S^1$ into two intervals, $S^1 = \mathcal I_1 \cup \mathcal I_2$. This splits $T^2$ along two cycles whose sum is trivial. 
One of the intervals represents the black hole system and the other one the gravitating bath.
The EE's associated with the split $S^1$ can be computed using minimal surfaces, as discussed above; they quantify the amount of information exchanged between the two subsystems.

For any split of the $S^1$ there are surfaces  stretching through the horizon of the $AdS_{d+1}$ black hole into the thermofield double.
Their area grows in time  (linearly at late times) \cite{Hartman:2013qma}.
These surfaces compete with `island surfaces', which cap off smoothly by connecting the two cycles along which the $T^2$ is split (fig.~\ref{fig:surfs-gen}). The latter are similar to island surfaces in the string theory versions of the brane world models \cite{Uhlemann:2021nhu}.
From the full $AdS_{d+1}\times T^2$ perspective there are no disconnected contributions to the entanglement wedge and no conflict with the gravitational dressing requirements discussed in \cite{Geng:2021hlu}. But from the perspective of the intermediate holographic description one system contributes to the EE of the other system through an island.
The area of the island surfaces is constant in time and limits the entropy growth.
Depending on the relative size of the intervals the island surfaces can either be dominant from the start, leading to a flat entropy curve, or become dominant after a certain amount of time, leading to a non-trivial Page curve.
We also find a second transition to a tiny island regime.
These phases with `critical' and `Page' values for the interval lengths precisely mirror the transitions in the braneworld models \cite{Geng:2020fxl} and their string theory uplifts \cite{Uhlemann:2021nhu}.
We discuss these phase transitions from the perspective of the IR solutions and in the RG flows.

In the last part we discuss the case $\Sigma=S^2$. The RG flow perspective allows to derive an EE interpretation for certain surfaces splitting the $S^2$ internal space in the IR geometry, and this case provides an instructive model for more general internal spaces in AdS/CFT. 
Surfaces splitting the internal space in general have to be anchored on an extremal sub-surface at the boundary of AdS \cite{Graham:2014iya}.
This constraint is respected by the split of $T^2$ discussed above.
It implies that an $S^2$ has to be split along an equatorial $S^1$. Instead of being able to pick an arbitrary region in the $S^2$, the boundary conditions determine how the equatorial $S^1$ is approached by the surface.

We find two scenarios, depending on the size of the $S^2$ relative to the AdS radius (set by the twist):
Fluctuations away from the equatorial $S^1$ at the boundary of AdS can have real or complex scaling dimensions, corresponding to power law behavior or damped oscillations. Both cases appear in AdS/CFT, for example in $AdS_3\times S^3\times S^3\times S^1$ \cite{Gukov:2004ym,Eberhardt:2017pty} where the size of the $S^3$'s can take a range of values so that both scenarios can be realized for surfaces splitting one of the $S^3$'s.
From the RG flows, where the $S^2$ is part of the field theory geometry in the UV, we obtain the following perspective: In both cases there are extremal surfaces anchored on the $S^2$ in the UV that reach into the IR solution.
But when the scaling dimensions in the IR solution are complex these surfaces are not minimal. Instead, actual minimal surfaces cap off in the UV before reaching the IR geometry. This leads to entanglement shadows, similar to those discussed in  \cite{Balasubramanian:2014sra,Balasubramanian:2017hgy}. An interpretation of minimal surfaces splitting the internal space in the IR solutions as EE is therefore not immediate.
When the scaling dimensions are real, on the other hand, the surfaces in the IR geometry can be obtained straightforwardly as limits of surfaces computing geometric EE's in the UV. These surfaces therefore have a clear interpretation as `internal space EE' in the IR geometry.

In summary, topologically twisted compactifications are interpreted in the spirit of wedge holography and used to model information transfer from a black hole to a gravitating bath. 
The resulting entropy curves reproduce the phenomenology in braneworld models. 
The setups are embedded in string theory and have concrete holographic duals in the form of topologically twisted compactifications of $\mathcal N=4$ SYM and 6d $\mathcal N=(2,0)$ theories.
From a broader perspective, the setups are a fruitful setting for tracing the transition from geometric EE's to non-geometric EE's, whose holographic study was initiated in \cite{Mollabashi:2014qfa,Karch:2014pma}. For twisted compactifications on $S^2$ we find that, depending on the size of the $S^2$, surfaces splitting the $S^2$ internal space can either directly inherit an EE interpretation from the RG flow perspective or lie in an entanglement shadow.

\medskip
{\bf Outline:}
In sec.~\ref{sec:wedge} the wedge holography interpretation of topologically twisted compactifications on $T^2$ is discussed in more detail. In sec.~\ref{sec:gen} we study EE's and Page curves in a general class of geometries which captures the IR fixed points of $T^2$ compactifications.
In sec.~\ref{sec:T2-flows} we discuss $\mathcal N=4$ SYM and 6d $\mathcal N=(2,0)$ theories on $T^2$ and study the EE's along the RG flows. Further examples on $T^2$ are discussed briefly. Compactifications on $S^2$ are discussed in sec.~\ref{sec:S2}.

\section{A twist on wedge holography}\label{sec:wedge}

\begin{figure}[h!]
	\subfigure[][]{
		\begin{tikzpicture}[scale=0.8]\label{fig:wedge-1}
			\draw (-2.5,0) -- (2.5,0);

			\draw[thick] (-0.8,0) -- (-2.5,-2/3*2.5);
			\draw[thick] (0.8,0) -- (2.5,-2/3*2.5);
			
			\draw [white,fill=gray,opacity=0.3] (-0.8,0) -- (-2.5,0) -- (-2.5,-2/3*2.5)--(-0.8,0);
			\draw [white,fill=gray,opacity=0.3] (0.8,0) -- (2.5,0) -- (2.5,-2/3*2.5)--(0.8,0);
			
			\node at (-0.6-0.8,-0.22) {\footnotesize $\theta_{1}$};
			\node at (0.6+0.8,-0.22) {\footnotesize $\theta_2$};
			\draw (-0.9-0.8,0) arc (180:225:25pt);
			\draw (0.9+0.8,0) arc (0:-45:25pt);	
			
			\node at (-3.4,0) {\footnotesize $z=0$};	
			\node at (-3.4,-1.7) {\footnotesize $z\rightarrow \infty$};	
			
			\node at (0.5,0.7) {\footnotesize $I\times \RR^{1,d-1}$};
			\node[rotate=-90] at (0,0.25) {$\Bigg\lbrace$};
		\end{tikzpicture}
	}
	\hskip 15mm
	\subfigure[][]{
		\begin{tikzpicture}[scale=0.8]\label{fig:wedge-2}
			
			\draw (-2.5,0) -- (2.5,0);

			\draw[thick] (0,0) -- (-1.7,-2/3*2.5);
			\draw[thick] (0,0) -- (1.7,-2/3*2.5);
			
			\draw [white,fill=gray,opacity=0.3] (0,0) -- (-2.5,0) -- (-2.5,-2/3*2.5) -- (-1.7,-2/3*2.5) --(0,0);
			\draw [white,fill=gray,opacity=0.3] (0,0) -- (2.5,0) -- (2.5,-2/3*2.5)-- (1.7,-2/3*2.5) -- (0,0);
			
			\node at (-0.6,-0.22) {\footnotesize $\theta_{1}$};
			\node at (0.6,-0.22) {\footnotesize $\theta_2$};
			\draw (-0.9,0) arc (180:225:25pt);
			\draw (0.9,0) arc (0:-45:25pt);
			
			\node at (-3.4,0) {\footnotesize $z=0$};	
			\node at (-3.4,-1.7) {\footnotesize $z\rightarrow \infty$};

			\node at (0.2,0.5) {\footnotesize $\RR^{1,d-1}$};
		\end{tikzpicture}
	}
	\caption{Wedge holography from CFT$_{d+1}$ on an interval. The $\RR^{1,d-1}$ part has been suppressed. The conformal boundary on the left is the product of an interval and $\RR^{1,d-1}$; on the right it is reduced to $\RR^{1,d-1}$.\label{fig:wedge}}
\end{figure}
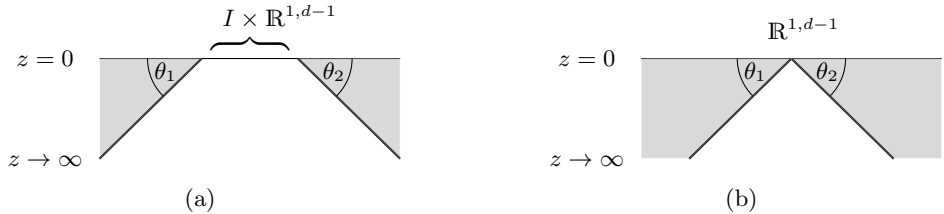

To elaborate on the codimension-3 wedge holography interpretation of topologically twisted compactifications on a Riemann surface $\Sigma$ we first recall the arguments of \cite{Akal:2020wfl} for codimension-2 wedge holography.
Starting point is $AdS_{d+2}$ cut off by two end-of-the-world (ETW) branes so that the remaining part of the conformal boundary contains an interval (fig.~\ref{fig:wedge-1}) and the dual is a CFT$_{d+1}$ on an interval.
The gravitational setup can be described by the metric
\begin{align}\label{eq:ds2-0}
	ds^2_{d+2}&=d\rho^2+e^{2\rho}(dx^2+ds^2_{\RR^{1,d-1}})~,
\end{align}
where $\rho$ is the $AdS_{d+1}$ radial coordinate. We will also use the conformal radial coordinate $z=e^{-\rho}$ for illustration. The ETW branes restricting the range of $x$ to $x\in(x_-,x_+)$ are located at 
\begin{align}
	x_\pm&=\pm a \pm b_\pm e^{-\rho}~.
\end{align}
The constants $b_\pm$ are related to the brane angles $\theta_{1/2}$ and fixed by the brane tensions.
This leaves a trapezoid region of $AdS_{d+2}$ parametrized by the $(x,\rho)$ coordinates, which may be described as an interval (the $x$ direction) warped over the radial coordinate (the vertical direction in fig.~\ref{fig:wedge-1}). 
The size of the interval in the CFT is set by $2a$.
The proper size of the interval in the geometry (\ref{eq:ds2-0}) diverges as $2a e^{\rho}$ in the UV, in accordance with the other field theory directions.

In the IR limit the CFT does not resolve the interval and reduces to a theory in one dimension less.
In the holographic dual this is seen from $x_\pm \approx b_\pm e^{-\rho}$ for $\rho\rightarrow -\infty$.
The proper length of the interval in the metric (\ref{eq:ds2-0}) approaches a constant $b_+-b_-$ in the IR.
From the IR perspective the geometry reduces to fig.~\ref{fig:wedge-2}. The interval reduces to a point at the conformal boundary, reducing the trapezoid to a wedge.
In the IR setup the interval has become the internal space. 
This can be made manifest by writing the $AdS_{d+2}$ metric in $AdS_{d+1}$ slicing, 
\begin{align}
	ds_{d+2}^2=d\tilde \rho^2+\cosh^2\!\tilde\rho\,ds^2_{AdS_{d+1}}~.
\end{align}
The two ETW branes limit $\tilde\rho$ to a finite range, the warp factor $\cosh^2\!\tilde\rho$ is bounded and the metric describes a warped product of $AdS_{d+1}$ over the interval parametrized by $\tilde\rho$.

For the topologically twisted compactifications the situation is analogous. 
We focus on $\Sigma=T^2$.
We keep the discussion general here and discuss concrete examples in sec.~\ref{sec:T2-flows}.
The UV solution is dual to a CFT$_{d+2}$ on $\RR^{d}\times T^2$. 
The metric for the full flow can be written as 
\begin{align}\label{eq:ds2-flow-gen}
	ds_{d+3}^2&=d\rho^2 + e^{2f(\rho)} ds^2_{\RR^{1,d-1}} + e^{2g(\rho)} ds^2_{T^2}~,
\end{align}
where 
\begin{align}
	f(\rho)&\sim \kappa_1 \rho~, & g(\rho) &\sim \kappa_1\rho~, & \text{for $\rho\rightarrow +\infty$}~,
	\nonumber\\
	f(\rho)&\sim \kappa_2 \rho~, & g(\rho) &\sim {\rm const}~, & \text{for $\rho\rightarrow -\infty$}~.
\end{align}
For $\kappa_1,\kappa_2>0$ the UV part of the geometry corresponds to $\rho\rightarrow \infty$ and the IR part to $\rho\rightarrow -\infty$.
In the UV the proper size of the torus diverges in accordance with the remaining field theory directions;
the conformal boundary of this asymptotically $AdS_{d+3}$ space is $\RR^d\times T^2$.
In the IR limit the size of the $T^2$ decouples from the $AdS$ scaling and approaches a constant. The torus becomes the internal space, while the radial direction $\rho$ and the $\RR^{1,d-1}$ slices form $AdS_{d+1}$.
In the IR the CFT reduces its dimension by two -- instead of a shrinking interval we have a shrinking $T^2$.

\begin{figure}
	\subfigure[][]{\label{fig:torus-wedge-1}
		\begin{tikzpicture}[scale=0.74]
			\node at (0,-3.4) {\includegraphics[width=0.19\linewidth]{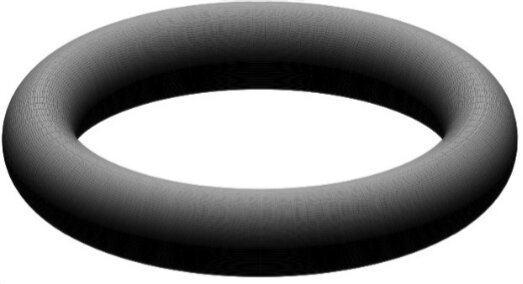}};
			\node at (0,-2.8) {\includegraphics[width=0.165\linewidth]{torus-2.png}};
			\node at (0,-2.2) {\includegraphics[width=0.14\linewidth]{torus-2.png}};
			\node at (0,-1.6) {\includegraphics[width=0.115\linewidth]{torus-2.png}};
			\node at (0,-1) {\includegraphics[width=0.09\linewidth]{torus-2.png}};
			
			\draw[dashed] (2.3,-3.5) -- (0.9,-0.4) -- (-0.9,-0.4) -- (-2.3,-3.5);
			
			\node at (2,-0.5) {\footnotesize $\rho=+\infty$};
			\node at (3.3,-3.4) {\footnotesize $\rho\rightarrow-\infty$};
			
			\node at (0.2,0) {\footnotesize $\RR^{1,d-1}\times T^2$};
		\end{tikzpicture}
	}
	\hskip 15mm
		\subfigure[][]{\label{fig:torus-wedge-2}
		\begin{tikzpicture}[scale=0.76]
			\node at (0,-3.5) {\includegraphics[width=0.17\linewidth]{torus-2.png}};						
			\node at (0,-3) {\includegraphics[width=0.14\linewidth]{torus-2.png}};			
			\node at (0,-2.5) {\includegraphics[width=0.11\linewidth]{torus-2.png}};
			\node at (0,-2) {\includegraphics[width=0.08\linewidth]{torus-2.png}};
			\node at (0,-1.5) {\includegraphics[width=0.05\linewidth]{torus-2.png}};
			\node at (0,-1) {\includegraphics[width=0.02\linewidth]{torus-2.png}};
			\draw[dashed] (2.1,-3.6) -- (0,-0.6) -- (-2.1,-3.6);
			
			\node at (1.2,-0.6) {\footnotesize $\rho=+\infty$};
			\node at (3,-3.5) {\footnotesize $\rho\rightarrow-\infty$};
			
			\node at (0.4,-0.2) {\footnotesize $\RR^{1,d-1}$};
		\end{tikzpicture}
	}
\caption{Wedge holography picture for $T^2$ compactification. On the left the torus is of finite size at the conformal boundary of $AdS$ (measured without the $e^{2\rho}$ factor in the metric). On the right it shrinks to zero size (again measured without the $e^{2\rho}$ factor) .\label{fig:torus-wedge}}
\end{figure}

The transition analogous to fig.~\ref{fig:wedge} for the torus compactification is illustrated schematically in fig.~\ref{fig:torus-wedge}.
In the IR the geometry simply becomes
\begin{align}\label{eq:ds2-AdSxT2}
	ds^2_{d+3}&=ds^2_{AdS_{d+1}} + ds^2_{T^2}~.
\end{align}
We transition from AdS$_{d+3}$/CFT$_{d+2}$ in the UV to AdS$_{d+1}$/CFT$_d$ in the IR.
In the language of \cite{Akal:2020wfl} this may be seen as codimension-3 holography for the generalized wedge consisting of the torus warped over the AdS$_{d+1}$ radial direction.\footnote{A proposal for (bottom-up) codimension-$n$ holography, based on CFTs with defects of higher codimension, can be found in \cite{Miao:2021ual}. Here we have RG flows with string theory embeddings and concrete CFT duals, and no defects.}
The benefit of this picture is that methods established for the AdS part of the geometry can be imported into the internal space in the IR setup. The example of interest here is the computation of EE's associated with splitting the torus, which can be started in the UV, where the standard Ryu/Takayanagi prescription can be applied, and then followed into the IR region where the torus becomes part of the internal space.\footnote{Geometric EE's on the torus were studied in \cite{Bueno:2016rma}. In contrast to the AdS soliton solutions studied there, which describe gapped states and close off in the IR, the setups considered here flow to non-trivial CFTs in the IR and correspondingly develop $AdS_{d+1}$ throats in asymptotically $AdS_{d+3}$ geometries. This  allows for the wedge holography picture and our main interest is in the IR geometry, where the torus is part of the internal space.}

\subsection{Splitting the internal space}

In the braneworld model a natural split into two subsystems can be obtained by considering the left and right ETW branes separately.
This has an analog in the topologically twisted compactifications as decomposing the $T^2$ internal space.

One way to motivate splitting the $T^2$ internal space
comes from the string theory uplifts of the braneworld models, in which a higher-dimensional internal space supersedes the wedge region with ETW branes. The string theory versions are based on Type IIB supergravity solutions in which the geometry is a warped product of $AdS_4$ and two $S^2$'s over a strip $\hat\Sigma$ \cite{DHoker:2007zhm,DHoker:2007hhe,Aharony:2011yc,Assel:2011xz}. On each boundary of the strip an $S^2$ collapses to form a closed internal space.
The decomposition into degrees of freedom represented by the two ETW branes corresponds to decomposing $\hat\Sigma$ \cite{Uhlemann:2021nhu}.
The solutions are dual to 3d $\mathcal N=4$ quiver gauge theories, where decomposing $\hat\Sigma$ should correspond to a decomposition of the quiver diagram. For the solutions used in sec.~4 of \cite{Uhlemann:2021nhu} the dual gauge theories were studied in \cite{Coccia:2020wtk}; the quivers have a reflection symmetry and the decomposition of $\hat\Sigma$ corresponds to a split of the quiver diagram into two halves (see also \cite{Bachas:2017rch,VanRaamsdonk:2021duo} for related discussions).

The features of the $T^2$ compactifications are analogous:
The geometry includes a smooth internal space instead of a wedge cut off by ETW branes.
The $T^2$ in the internal space represents a $U(1)\times U(1)$ global symmetry in the dual CFT.
Decomposing $T^2$ should amount to decomposing the CFT according to this $U(1)\times U(1)$ symmetry (instead of decomposing a quiver diagram). EE's associated with $U(1)$ symmetries were studied for example in \cite{Taylor:2015kda}.
We focus on deompositions of the torus in which one $S^1$ is decomposed as $S^1=\mathcal I_1\cup \mathcal I_2$, such that\footnote{An analog in the context of 3d quiver gauge theories might be to start from a circular quiver, for which holographic duals were discussed in \cite{Assel:2012cj}, and cut the quiver diagram twice.}
\begin{align}\label{eq:T2-decomp}
	T^2\,&=\,\mathcal I_1\times S^1 \, \cup \,  \mathcal I_2\times S^1~.
\end{align}
Analogs of the brane angles can be identified as follows:
The brane angles are fixed by the tensions of the ETW branes, which are a measure for the number of degrees of freedom they represent.
This can be phrased in terms of the size of the internal space.
In fig.~\ref{fig:wedge-2} the $x$-interval corresponds to the internal space and has proper length $b_+-b_-$. 
The left and right ETW branes account for the $(x_-,0)$ and $(0,x_+)$ parts of the interval, respectively.	
The volume of the internal space is set by the sum of the brane tensions, while the relative size of the parts ascribed to the left and right subsystems is set by the ratio of the brane tensions.
For the $T^2$ compactification the size of the torus relative to the AdS curvature counts the total number of degrees of freedom and corresponds to the sum of the brane tensions, while the relative volume of the two parts into which the torus is decomposed corresponds to the relative magnitude of the brane tensions.
Schematically,
\begin{align}
	|\mathcal I_{1/2}| & \ \longleftrightarrow \ \cot\theta_{1/2}~.
\end{align}

An `intermediate' holographic description can be obtained in the braneworld models by separately geometrizing the degrees of freedom represented by the two ETW branes. This leads to two gravitating systems with asymptotically-AdS boundary conditions which interact.
A natural analog in the $T^2$ compactifications can be obtained by geometrizing the degrees of freedom represented by the 	two parts of the torus decomposition in (\ref{eq:T2-decomp}) separately, resulting similarly in two gravitating systems with AdS boundary conditions which interact.

In the setup in fig.~\ref{fig:wedge-1} the energy-momentum tensors associated with the $d$-dimensional defect degrees of freedom on the left and right are not conserved, due to interactions with the ambient CFT$_{d+1}$. As a result the $(d+1)$-dimensional graviton localized near the ETW brane is massive. In  fig.~\ref{fig:wedge-2}, on the other hand, there is a conserved $d$-dimensional energy-momentum tensor and a massless $(d+1)$-dimensional graviton. But when the system is decomposed into left and right sectors neither sector has access to the massless mode. This avoids the constraints discussed in \cite{Geng:2021hlu} and allows for the formation of islands. 
The decomposition of $T^2$ in fig.~\ref{fig:torus-wedge-2} is in line with that picture: 
The total energy-momentum tensor of the CFT$_d$ is conserved and the dual has a massless $(d+1)$-dimensional graviton, which corresponds to a flat profile on $T^2$. But decomposing the torus leads to two systems which interact, and the individual systems do not have separately conserved energy-momentum tensors. 
In the intermediate holographic description neither of the gravity duals for the two subsystems has access to the massless graviton. 
This again avoids the constraints of \cite{Geng:2021hlu}.
Similar remarks apply for the string theory versions of the model in fig.~\ref{fig:wedge-2}, which have a massless graviton mode with uniform support on the strip $\hat\Sigma$ but when subsystems are defined by decomposing $\hat\Sigma$ each sector only has access to massive gravitons.

\subsection{Black holes and baths}

To model communicating black holes we generalize the solutions discussed above to incorporate finite temperature from the CFT$_d$ perspective.
We start from the IR solution (\ref{eq:ds2-AdSxT2}), which in the examples to be discussed below describes a supersymmetric compactification of a CFT$_{d+3}$ on $T^2$. Replacing the AdS$_{d+1}$ factor in the geometry by a planar AdS$_{d+1}$ black hole leads to a non-supersymmetric configuration which still solves the equations of motion.
We use the metric
\begin{align}\label{eq:ds2-bh-d}
	ds^2_{d+1}&=\frac{dr^2}{b(r)}+e^{2r}\left(-b(r)dt^2+d\vec{x}^2\right),&b(r)&=1-e^{-d(r-r_h)}~.
\end{align}
The horizon is at $r=r_h$ and the conformal boundary at $r=\infty$.
In CFT$_{d+2}$ terms this replacement introduces a temperature which is small compared to the energy scale associated with the compactification.
In the tortoise coordinate $u=\frac{2}{d}\tanh^{-1}\sqrt{b(r)}$, the black hole metric becomes
\begin{align}\label{eq:tortoise}
	ds^2_{d+1}&=du^2+e^{2r_h}\cosh^{4/d}\left(\frac{ud}{2}\right)\left[-\tanh^2\left(\frac{ud}{2}\right)dt^2+d\vec{x}^2\right].
\end{align}
The original exterior region corresponds to $u>0$, the second exterior region in the thermofield double to $u<0$.
The ER bridge connecting the two regions corresponds to a contour in the complex $u$ plane.

One of the subsystems associated with a decomposition of the torus can now be designated as black hole and the other as bath. Gravity is dynamical in both systems and the bath is gravitating.
The information exchanged between the two systems can be quantified by the EE associated with the split of the $T^2$ internal space.

\section{Entanglement entropy and Page curves}\label{sec:gen}

In this section we work with a general metric describing a (possibly warped) product of AdS$_{d+1}\times S^1$ over an internal space and discuss EE's associated with decomposing the $S^1$.
The setup covers examples like $\mathcal N=4$ SYM and 6d $\mathcal N=(2,0)$ theories compactified on $T^2$, to be discussed in sec.~\ref{sec:T2-flows}, but is more general. Motivated by the discussion in the previous section we use the Ryu/Takayanagi prescription to compute EE's associated with splitting the internal $S^1$.
The metric takes the form
\begin{align}\label{eq:ds2-gen}
&&	ds^2_D&=f^2\left(ds^2_{{d+1}}+d\phi^2\right)+ds^2_{\mathcal M_{D-d-2}}~, & \phi \sim \phi+L
\end{align}
where 
$ds^2_{d+1}$ is a unit-radius AdS$_{d+1}$ metric and
$\mathcal M_{D-d-2}$ is an internal space of approriate dimension to obtain a $D$-dimensional solution. $f$ is a function on $\mathcal M_{D-d-2}$ describing a possibly warped product of $AdS_{d+1}\times S^1$ over $\mathcal M_{D-d-2}$ and  $\phi$ parametrizes an $S^1$.
The metric (\ref{eq:ds2-AdSxT2}) corresponds to $D=d+3$, $f=1$ and $ds^2_{\mathcal M_{D-d-2}}=ds^2_{S^1}$.
The full solution (\ref{eq:ds2-gen}) may have other non-trivial fields, which are all assumed to respect the $AdS_{d+1}$ isometries.
We can then replace $AdS_{d+1}$ by a different Einstein space with negative curvature and still obtain a solution to the equations of motion.
In particular, we can use the black hole metric (\ref{eq:ds2-bh-d}).

To interpret the system as a black hole coupled to a bath we decompose the $S^1$ parametrized by $\phi$, which is assumed to have length $L$ with $\phi\in (-L/2,L/2)$. We then decompose
\begin{align}\label{eq:split}
	S^1_\phi&=\mathcal I_1 \cup \mathcal I_2~, &
	\mathcal I_1 &= (-\phi_0,\phi_0)~, & \mathcal I_2 = S^1_\phi\setminus \mathcal I_1~.
\end{align}
Following the discussion in sec.~\ref{sec:wedge}, the lengths of the intervals,
\begin{align}
	\ell_1&=|\mathcal I_1|=2\phi_0~, & \ell_2&=|\mathcal I_2|=L-2\phi_0~,
\end{align}
correspond to the brane angles in the braneworld models.
We designate the system represented by $\mathcal I_1$ as the black hole system and the system represented by $\mathcal I_2$ as bath. Gravity is dynamical in both sectors.
To realize a black hole coupled to a weakly-gravitating bath, the $S^1$ should be split such that the size of the interval $\mathcal I_1$ representing the black hole system is small compared to the size of the interval $\mathcal I_2$ representing the bath, 
\begin{align}\label{eq:l1l2}
	\ell_1<\ell_2~.
\end{align}
The central charge associated with the $\mathcal I_2$ system is then larger than that of the $\mathcal I_1$ system and the gravity dual of the $\mathcal I_2$ system has a smaller Newton constant. This corresponds in the braneworld models to the black hole brane being at a larger angle than the bath brane.

The geometry (\ref{eq:ds2-gen}) is assumed to arise as IR fixed point of an RG flow where the $S^1$ is part of the field theory geometry in the UV. This allows us to compute EE's associated with the internal space decomposition (\ref{eq:split}) using a generalization of the Ryu/Takayanagi prescription.\footnote{
Surfaces splitting the internal space of an asymptotically-AdS geometry have to be anchored at the boundary of AdS on a sub-surface which is itself extremal \cite{Graham:2014iya}.
Any choice of two points on $S^1$ constitutes an extremal sub-surface (and so does the union of two $S^1$ cycles in $T^2$). So there are no additional constraints.}

\subsection{Extremal surfaces}

\begin{figure}
	\begin{tikzpicture}
		
		\draw[green,very thick] (-0.05,1.5) -- (-0.05,-0.5);
		\draw[green,very thick,dashed] (0.05,2.5) -- (0.05,0.5);
		
		\draw[dashed, thick] (0,0) ellipse (1.0cm and 0.5cm);
		\draw[thick] (0,2) ellipse (1.0cm and 0.5cm);
		\draw[thick] (-1,0) -- (-1,2);
		\draw[thick] (1,0) -- (1,2);

		\draw[blue,thick] (-0.05,1.5) .. controls (-0.05,1.3) and (-0.4,0.5) .. (-0.99,0.6);
		\draw[blue,thick,dashed] (0.05,2.5) .. controls (0.05,2.3) and (-0.4,0.64) .. (-0.99,0.62);
		
		\node at (0.8,2.55) {\footnotesize $\mathcal I_2$};
		\node at (-0.8,2.55) {\footnotesize $\mathcal I_1$};
		
		\draw[->] (1.2,0.8) -- (1.2,1.2);
		\node at (1.4,1) {\small $r$};
		\node at (1.3,0) {\small $r_h$};
		\node at (0,-0.75) {\small $\phi$};
		
		\draw[white] (0,-1.1) circle (1pt);

	\end{tikzpicture}
	
	\caption{Schematic illustration of island and HM surfaces. The black dashed line at the bottom is the horizon, the top the conformal boundary. The region between the green and blue curves is the island.\label{fig:surfs-gen}}
\end{figure}
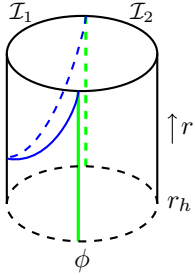

The EE associated with the split (\ref{eq:split}) is determined by the competition between two types of extremal surfaces, illustrated in fig.~\ref{fig:surfs-gen}. We refer to surfaces stretching through the horizon into the thermofield double as HM surfaces  \cite{Hartman:2013qma}.
These surfaces wrap  $\mathcal M_{D-d-2}$ and are anchored at two points on $S^1$ at the same time in the two asymptotic regions of the extended $AdS_{d+1}$ black hole geometry (evolving forward in time in both regions).
Their evolution in $AdS_{d+1}$ is time dependent.
The angular dependence of these surfaces simply amounts to fixing
\begin{align}
	\phi&=\pm \phi_0~.
\end{align}
Such surfaces are extremal and they exist for all values of $\phi_0$. 
At a point of time reflection symmetry, which can be chosen as $t=0$, they wrap a constant time slice of the $AdS_{d+1}$ black hole, and their area is given by twice the area in one of the regions. 
In the coordinates (\ref{eq:ds2-bh-d}), the area in one exterior region is
\begin{align}\label{eq:area-HM}
	A_{{\rm HM},t=0}&=C\int_{r_h}^{r_\epsilon} dr\,\frac{e^{(d-1)r}}{\sqrt{b(r)}}
	=\frac{C}{d-1}\left[\frac{1}{\epsilon ^{d-1}}- e^{(d-1)r_h}\frac{\sqrt{\pi } \Gamma \left(\frac{1}{d}\right)}{\Gamma
		\left(\frac{1}{d}-\frac{1}{2}\right)}\right]~,
\end{align}
where a cut-off $r_\epsilon=-\ln \epsilon$ with $\epsilon\ll 1$ was used for the $AdS$ radial coordinate to obtain the second equality. The coefficient $C$ is given by
\begin{align}\label{eq:C-def}
	C&=2V_{\RR^{d-1}}\int_{\mathcal M_{D-d-2}} f^{d+1}\vol_{\mathcal M_{D-d-2}}~,
\end{align}
where the factor $2$ accounts for the two branches of the surface at $\phi=\pm \phi_0$.
For generic times the surfaces wrap the $u>0$ and $u<0$ regions in the coordinates (\ref{eq:tortoise}), and a contour along the imaginary axis which describes the ER bridge between the two asymptotic regions. The resulting area grows in time. The explicit time dependence can be calculated analogously to app.~A of \cite{Geng:2020fxl} and is linear at late times.
If the HM surface were the only surface available this would indicate an unbounded growth of the EE associated with the split (\ref{eq:split}) and constitute an information paradox as discussed in \cite{Almheiri:2019yqk}.

The actual time evolution of the EE is determined by the competition between HM surfaces and surfaces that do not stretch through the horizon. Such surfaces extend along a constant-time slice of the $AdS_{d+1}$ black hole geometry, wrap $\mathcal M_{D-d-2}$ and are anchored at the conformal boundary at the same two points $\phi=\pm \phi_0$ on $S^1$. Instead of reaching through the horizon they cap off smoothly in each exterior region separately, as shown in fig.~\ref{fig:surfs-gen}.
These surfaces can be described (in each exterior region) by two branches of a single embedding function 
\begin{align}
	\phi&=\pm \phi(r)~, & \lim_{r\rightarrow \infty}\phi(r)&=\phi_0~.
\end{align}
The two branches meet at a cap-off point $r_\star$. The surface can close to either side on the torus: the two branches can meet either at $\phi=0$ or at the antipodal point $\phi=L/2$.
We focus without loss of generality on surfaces closing at $\phi=0$; the other case can be obtained by shifting $\phi_0\rightarrow L/2-\phi_0$.
So at the cap-off point,
\begin{align}
	\phi(r_\star)&=0~, & 
	\phi'(r)&\sim 1/\sqrt{r-r_\star}~.
\end{align}
The area of such a surface is given by
\begin{align}\label{eq:area-gen}
	A&=C\int_{r_\star}^{r_\epsilon} dr\,e^{(d-1)r}\sqrt{\frac{1}{b(r)}+{\phi^\prime}^2}~,
\end{align}
with the overall coefficient $C$ as defined in (\ref{eq:C-def}) where the factor two again accounts for the two branches.
Since $\phi$ appears in $A$ only through its derivative a first integral for the equation of motion can be obtained straightforwardly.
For $d>1$ ($d=1$ will be discussed in sec.~\ref{sec:ex-contd}),
\begin{align}\label{eq:dphi}
	\phi^\prime&=\frac{1}{\sqrt{b(r)\left(e^{2(d-1)(r-r_\star)}-1\right)}}~.
\end{align}
The derivative diverges at $r\rightarrow r_\star$, where the surface caps off. The solution $\phi$ is obtained by integrating (\ref{eq:dphi}) with the initial condition $\phi(r_\star)=0$.
The turning point $r_\star$ has to be determined so that the surface is anchored at $\pm \phi_0$ at $r\rightarrow\infty$.
The area is constant in time and given by 
\begin{align}\label{eq:area-island-2}
	A&=C\int_{r_\star}^{r_\epsilon} dr\,e^{(d-1)r}\sqrt{\frac{\coth\left((d-1)(r-r_\star)\right)+1}{2b(r)}}~.
\end{align}

As discussed around (\ref{eq:l1l2}), to realize a black hole coupled to a weakly-gravitating bath, the size of the interval $\mathcal I_2$ should be large compared to the size of $\mathcal I_1$. The surfaces capping off at $\phi=0$ then dominate those capping off at $\phi=L/2$.  The former are island surfaces in the following sense: In the UV they are anchored at $\phi=\pm \phi_0$ and separate the black hole system from the bath. But in the region $r<r_\star$ they capture the whole $S^1$. So degrees of freedom from the $\mathcal I_1$  black hole system contribute to the EE of the $\mathcal I_2$ bath system. The part of the $\mathcal I_1$ region between the minimal surface and the horizon (extending through the interior of the black hole into the second exterior region and bounded there by the second copy of the island surface), is the island.

\subsection{Critical and Page lengths}

The competition between island and HM surfaces determines whether a Page curve emerges or if the entropy curve is flat.
We start the discussion at zero temperature, i.e.\ with $b(r)=1$, since it gives clean access to some of the critical parameters. 
At zero temperature (\ref{eq:dphi}) can be integrated in closed form. With the initial condition $\phi(r_\star)=0$ we find, for $d>1$,
\begin{align}
	\phi(r)&=\frac{1}{d-1} \tan ^{-1}\sqrt{e^{2 (d-1) (r-r_\star)}-1}~.
\end{align}
The limit value at the conformal boundary is independent of $r_\star$ as a result of conformal invariance.\footnote{If $\phi$ were part of the field theory directions it would scale with the field theory coordinates under conformal transformations. But here it is part of the internal space and as such does not scale.}
We denote the asymptotic value of $\phi$ in the zero-temperature background as {\it critical value}
\begin{align}\label{eq:phi0-crit}
\phi_{0,{\rm crit}}\equiv\lim_{r\rightarrow\infty}\phi(r)&=\frac{\pi}{2(d-1)}~.
\end{align}
Surfaces which cap off before reaching the Poinca\'e horizon and are not self-intersecting thus only exist for $L> L_{\rm crit}$ with
\begin{align}\label{eq:L-bound}
	L_{\rm crit}= \frac{\pi}{d-1}~.
\end{align}
For smaller $L$  surfaces which cap off at finite $r_\star$ wind around the $S^1$ and self-intersect. For $L$ larger than (\ref{eq:L-bound}), island surfaces exist for $|\mathcal I_1|= 2\phi_{0,{\rm crit}}$ or  $|\mathcal I_2|= 2\phi_{0,{\rm crit}}$.
For other splits of the $S^1$ only the HM surface is available. 
The area of the HM surface is constant in time at zero temperature and there is no information paradox.
The difference between the area of the island surface in (\ref{eq:area-island-2}) and the area of the HM surface (obtained for $r_\star\rightarrow -\infty$), is independent of $r_\star$ and vanishes.

\begin{figure}
	\subfigure[][]{\label{fig:surfs-T}
		\includegraphics[width=0.4\linewidth]{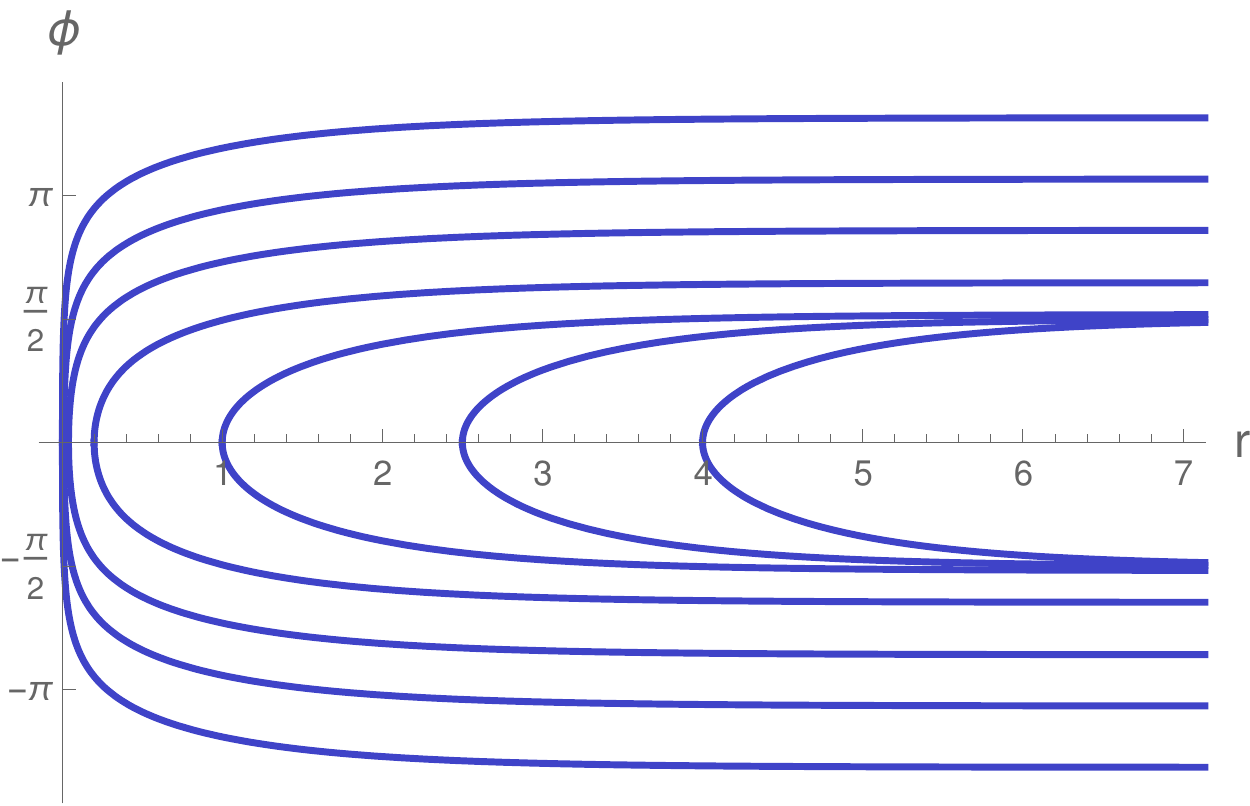}
	}
	\hskip 10mm
	\subfigure[][]{\label{fig:anchor-area-a}
		\includegraphics[width=0.42\linewidth]{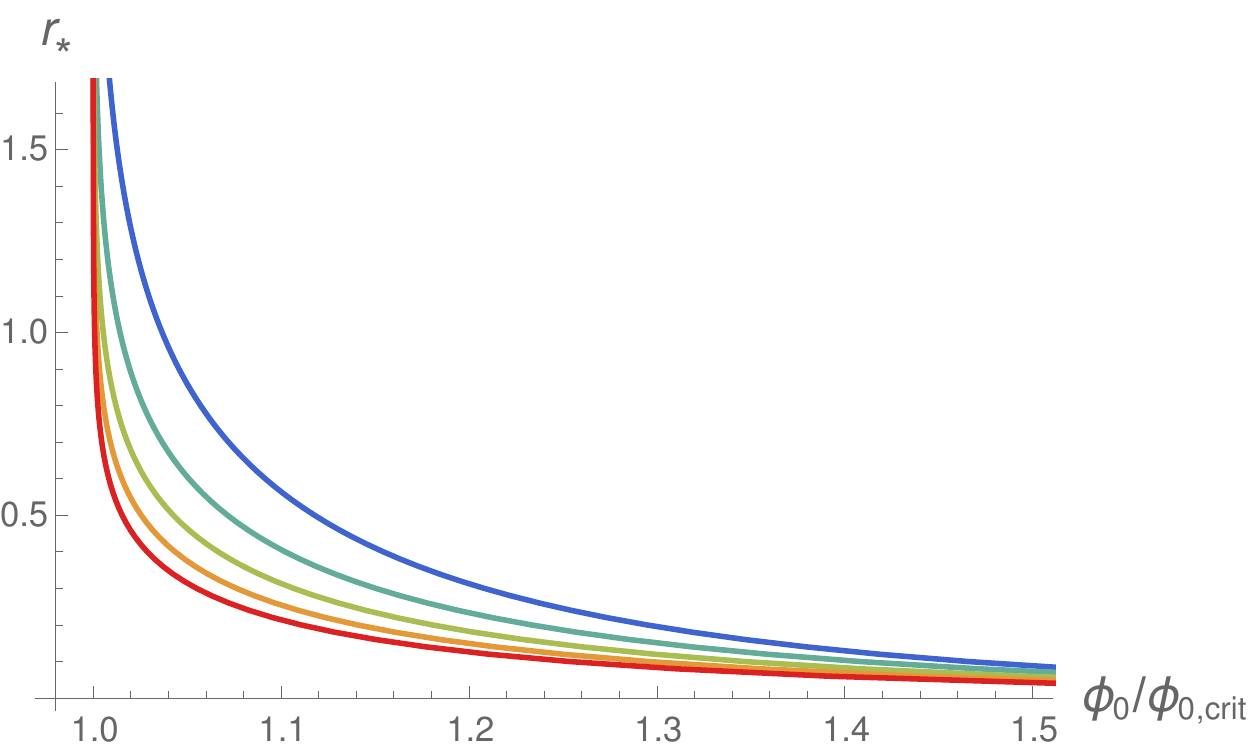}
	}
	\caption{Left: Island surfaces for $d=2$ with $r_h=0$. The surfaces reach beyond $\phi_{0,{\rm crit}}=\pi/2$. As $r_\star$ moves to the UV, the surfaces approach the zero temperature surface. Right: Turning point $r_\star$ as function of $\phi_0$ with $r_h=0$; the curves are for $d=2,3,4,5,6$ from bottom to top. The turning point diverges for $\phi_0\rightarrow \phi_{0,{\rm crit}}$.
		\label{fig:surfs-anchor}}
	
\end{figure}

For finite temperature (\ref{eq:dphi}) can be integrated numerically. For $d>1$ island surfaces exist for $\phi_0\geq\phi_{0,{\rm crit}}$. They flatten out at the horizon for $r_\star\rightarrow r_h$ and cover more of the $S^1$ than at zero temperature.
This is shown in fig.~\ref{fig:surfs-T}.
The cap-off points $r_\star$ approach the horizon for $\phi_0\rightarrow\infty$ and diverge towards the conformal boundary for $\phi_0\rightarrow\phi_{0,{\rm crit}}$ (fig.~\ref{fig:anchor-area-a}).

\begin{figure}
	\includegraphics[width=0.45\linewidth]{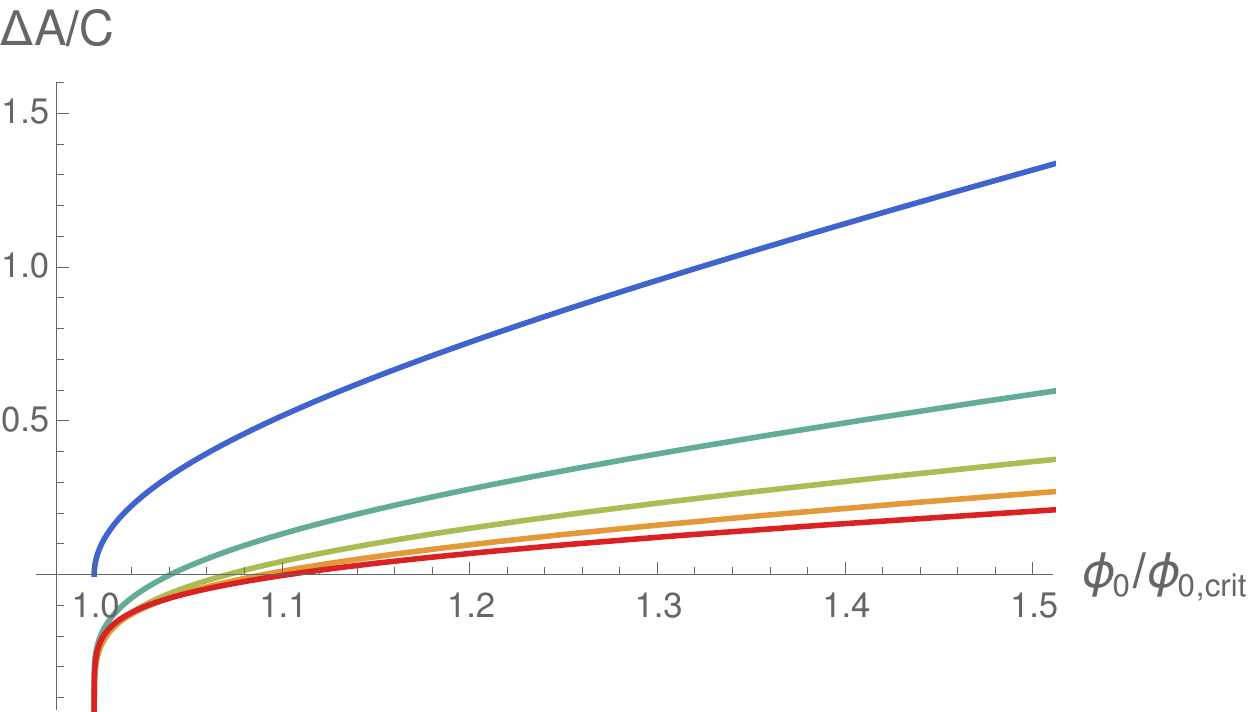}
	\caption{Area difference between island and HM surfaces at $t=0$, $\Delta A = A_{\rm island}-A^{t=0}_{\rm HM}$, normalized to $C$ in (\ref{eq:C-def}), from top to bottom for $d=2,3,4,5,6$.
		\label{fig:area}}
\end{figure}

The form of the entropy curve is determined by the competition between island and HM surfaces.
The area differences at $t=0$ are shown in fig.~\ref{fig:area}.
The area differences at the critical value $\phi_{0,{\rm crit}}$ can be obtained analytically, by evaluating the integral (\ref{eq:area-island-2}) for $r_\star\rightarrow \infty$. We find
\begin{align}
	\Delta A\equiv A_{\rm island}-A^{t=0}_{\rm HM}\,\big\vert_{\phi_0=\phi_{0,{\rm crit}}}&=C e^{(d-1)r_h} \frac{\sqrt{\pi}\,\Gamma\left(\frac{1}{d}\right)}{(d-1)\Gamma\left(\frac{1}{d}-\frac{1}{2}\right)}~.
\end{align}
For $AdS_3$ with $d=2$ the area difference at $\phi_{0,{\rm crit}}$ vanishes. 
For generic $d>2$ the area difference is negative for $\phi_0$ near the critical value. The Page curve for $\phi_{0}=\phi_{0,{\rm crit}}$ is flat in both cases.
For $d=2$ the entropy curve is non-trivial for any $\phi_0>\phi_{0,{\rm crit}}$.
For $d>2$ there are distinct {\it Page values} $\phi_{0,P}$ where a transition occurs from a flat entropy curve to a non-trivial Page curve.
Numerically,
\begin{align}\label{eq:phi-P}
	d=3:& \quad \phi_{0,P}\approx 1.041\,\phi_{0,{\rm crit}}~,
&
	d=4:& \quad \phi_{0,P}\approx 1.071\,\phi_{0,{\rm crit}}~,
	\nonumber\\
	d=5:& \quad \phi_{0,P}\approx 1.091\,\phi_{0,{\rm crit}}~,
&
	d=6:& \quad \phi_{0,P}\approx 1.104\,\phi_{0,{\rm crit}}~.
\end{align}
For $\phi_0>\phi_{0,P}$  the entropy follows non-trivial Page curves, below $\phi_{0,P}$ the entropies are constant.
The fact that the anchor point $r_\star$ approaches the conformal boundary when $\phi_0\rightarrow \phi_{0,{\rm crit}}$ suggests a `tiny island' regime, similar to the findings in the braneworld models \cite{Geng:2020fxl} and their string theory uplifts \cite{Uhlemann:2021nhu}. This will be discussed from the RG flow perspective in the next section.

To summarize, for small black hole systems with $\phi_0 < \phi_{0,{\rm crit}}$ we find a tiny island regime; for a small range of intermediate black hole systems with $\phi_{0,{\rm crit}}<\phi_0<\phi_{0,P}$ we find actual island surfaces but still flat entropy curves, while for large enough black hole systems with $\phi_0>\phi_{0,P}$ we find non-trivial Page curves.
The results are compatible with the discussions of \cite{Laddha:2020kvp,Raju:2020smc,Raju:2021lwh}, which argued (for asymptotically flat spaces) that geometrically defining a radiation system with dynamical gravity leads to flat entropy curves (for alternative discussions see \cite{Krishnan:2020oun,Krishnan:2020fer,Ghosh:2021axl}). Here the radiation system is defined according to internal quantum numbers and in that sense `non-geometrical'.

The phase transitions between tiny island regime, flat entropy curve and Page curve precisely parallel the discussions in \cite{Geng:2020fxl,Uhlemann:2021nhu}.
In the Karch/Randall models the brane angles of the two ETW branes, which determine the number of degrees of freedom represented by each brane, can be chosen independently. Here this corresponds to the lengths of the two intervals, $\ell_1$ and $\ell_2$.
If both are below the critical length there are no regular island surfaces capping off to either side.
If one of them is below and the other above the critical length, the EE is determined by the tiny island closing to the shorter side and the entropy curve is flat. Only if both are above the Page length do we get a non-trivial Page curve.
These cases correspond for the two brane angles in the Karch/Randall models to neither, one, or both exceeding the critical/Page values.

The EE discussion and metric in (\ref{eq:ds2-gen}) in principle cover more general backgrounds than twisted compactifications on $T^2$. 
For example the $AdS_3\times S^3\times T^4$ \cite{David:2002wn} and $AdS_3\times S^3\times S^3\times S^1$ solutions in Type IIB \cite{Gukov:2004ym,Eberhardt:2017pty}.
Another interesting example are the J-fold solutions of \cite{Assel:2018vtq,Bobev:2019jbi}.
The solutions in \cite{Bobev:2019jbi} are constructed in 5d supergravity and the metric is a product $AdS_4\times S^1$.
The dilaton has non-trivial dependence on the $S^1$ and is periodic only up to an $SL(2,\RR)$ transformation, but it does not enter EE computations. The solutions can be uplifted to 10d and are  dual to 3d $\mathcal N=1$ theories.
The advantage of the topologically twisted compactifications is the RG flow perspective, in which the IR internal space originates from the CFT geometry in the UV. 
The above discussion applies more generally if the internal space EE can be made sense of independently.

\section{From geometric to non-geometric entropy} \label{sec:T2-flows} 

In this section concrete examples of topologically twisted compactifications on $T^2$ will be discussed, and we discuss the behavior of the EE's along the RG flows where they correspond to geometrically defined subsectors.\footnote{%
The impact of relevant deformations on Page curves was discussed in braneworld models in \cite{Caceres:2021fuw}. Here we focus on compactifications of CFTs as a way to associate EE with surfaces splitting the internal space.}
We focus on $\mathcal N=4$ SYM and 6d $\mathcal N=(2,0)$ theories on $T^2$. 5d SCFTs on $T^2$ and compactifications to $AdS_2$ are discussed briefly. This covers $AdS_{d+1}$ IR fixed points with $d=1,2,3,4$.
We focus on flows which are supersymmetric at zero temperature, which ensures stability of the IR fixed points.

\subsection{\texorpdfstring{Flows to AdS$_3$: $\mathcal N=4$ SYM on $T^2$}{N=4 SYM on T**2}}

We start with $\mathcal N=4$ SYM on $T^2$, leading to  $AdS_3\times T^2$ in the IR. This is an instructive starting point and the higher-dimensional cases follow analogously. 
Supergravity solutions describing twisted compactifications of $\mathcal N=4$ SYM on a Riemann surface $\Sigma$ with constant curvature $R=2\kappa$, with $\kappa=1,0,-1$ for $S^2$, $T^2$ and hyperbolic surfaces, respectively, were discussed in \cite{Benini:2013cda}. Solutions for $\Sigma=T^2$ were first constructed in \cite{Almuhairi:2011ws,Donos:2011pn}; generalizations to 4d $\mathcal N=1$ theories can be found in \cite{Benini:2015bwz}.
The twists are characterized by a choice of $SO(6)$ R-symmetry background $A_\mu$ such that $F\propto a_I T^I \vol_\Sigma$, where $T^I$ with $I=1,2,3$ are generators of an $SO(2)^3$ subgroup of $SO(6)$ and $a_I$ are parameters specifying the twist, with $a_1+a_2+a_3=-\kappa$.

The supergravity duals are constructed in the STU model, which is a consistent truncation of 5d maximal gauged supergravity, which in turn is a consistent truncation of Type IIB supergravity on $S^5$. The field content of the STU model comprises the metric, three Abelian gauge fields $A^I$ and two real scalars $\phi_1$, $\phi_2$.
Everything we need here is captured in the STU model; the minimal surfaces we discuss wrap the entire internal space in the uplift to Type IIB.

The $AdS_3\times \Sigma$ solution dual to the 2d SCFT arising as IR fixed point of $\mathcal N=4$ SYM on $\Sigma$ takes the form
\begin{align}\label{eq:ds2-AdS3N4}
	ds^2&=e^{2f_0}ds^2_{AdS_3}+e^{2g_0}ds^2_{\Sigma}~,
	&
	F^I&=-a_I \vol_{\Sigma}~,
\end{align}
The scalar fields are constant and given by
\begin{align}\label{eq:N4SYM-IR1}
	e^{\sqrt{6}\phi_1}&=\frac{a_3^2(a_1+a_2-a_3)^2}{a_1a_2(-a_1+a_2+a_3)(a_1-a_2+a_3)}~,
	&
	e^{\sqrt{2}\phi_2}&=\frac{a_2(a_1-a_2+a_3)}{a_1(-a_1+a_2+a_3)}~,
\end{align}
with 
\begin{align}
	e^{6g_0}&=\frac{a_1^2a_2^2a_3^2}{\Pi}~, & 	\Theta&=a_1^2+a_2^2+a_3^2-2(a_1a_2+a_1a_3+a_2a_3)~,
	\nonumber\\
	e^{3f_0}&=-\frac{8a_1a_2a_3\Pi}{\Theta^3}~,
	&
	\Pi&=(-a_1+a_2+a_3)(a_1-a_2+a_3)(a_1+a_2-a_3)~.
	\label{eq:ThetaPi}
\end{align}
We focus here on the $T^2$ compactification. 
The quantization conditions for the $a_I$ and the volume as given in (3.6), (3.7) of \cite{Benini:2013cda} are
\begin{align}\label{eq:N4SYM-T2-norm}
	a_I&\in\ZZ~, & \int\vol_{\Sigma_{\mathfrak{g}=1}}&=2\pi~.
\end{align}
The allowed $a_I$ for a well-behaved supergravity solution to exist are discussed in appendix $F$ of \cite{Benini:2013cda}. The result is that two $a_I$ have to be strictly positive, which implies that the third is negative.
For the maximally supersymmetric $T^2$ compactification with $a_I=0$ the flow is singular in the IR.

We start with the $AdS_3$ IR fixed point solution and introduce a black hole/finite temperature.
Replacing the $AdS_3$ part in (\ref{eq:ds2-AdS3N4}) by the black hole metric (\ref{eq:ds2-bh-d}) with $d=2$ leads to the dual of the 2d CFT at finite temperature.
This amounts to compactifying $\mathcal N=4$ SYM on $\Sigma$ with a twist, flowing to the IR fixed point to obtain a 2d CFT, and then considering the 2d CFT at a temperature which is small compared to the compactification scale.
In supergravity terms the black hole solution is asymptotic to $AdS_3\times \Sigma$, which itself arises as IR fixed point of a solution with $AdS_5$ UV asymptotics in 5d supergravity and can be uplifted to Type IIB.

The torus is the internal space and the Page curve discussion of sec.~\ref{sec:gen} applies.
The metric (\ref{eq:ds2-AdS3N4}) with the torus volume as in (\ref{eq:N4SYM-T2-norm}) can be rewritten to match the convention of sec.~\ref{sec:gen} as
\begin{align}\label{eq:ds2-AdS3T2}
	&&	ds^2&=e^{2f_0}\left(ds^2_{AdS_3}+d\phi^2+d\chi^2\right)~,
	& \phi&\sim \phi+L~,
	\nonumber\\
	&&	&&\chi&\sim\chi+\frac{2\pi}{L}e^{2g_0-2f_0}~.
\end{align}
The volume of the torus relative to the $AdS_3$ radius is set by $e^{2g_0-2f_0}$. For $a_1+a_2+a_3=0$,
\begin{align}\label{eq:T2-size-N4SYM}
	e^{2g_0-2f_0}=\frac{\Theta^2}{4\Pi}&=\frac{\left(a_1^2+a_1a_2+a_2^2\right)^2}{2a_1a_2(a_1+a_2)}~.
\end{align} 
If the $S^1$ parametrized by $\phi$ is decomposed, with $\chi$ parametrizing the internal space $M_{D-d-2}$ in (\ref{eq:ds2-gen}), the results of sec.~\ref{sec:gen} can be applied directly, e.g.\ with the critical $L$ for finding non-trivial Page curves given in (\ref{eq:L-bound}).
By appropriately choosing the volume and shape of the torus and the decomposition of $S^1_\phi$, the scenarios with flat or non-trivial entropy curves discussed in sec.~\ref{sec:gen} can all be realized.

\begin{figure}
	\includegraphics[width=0.24\linewidth]{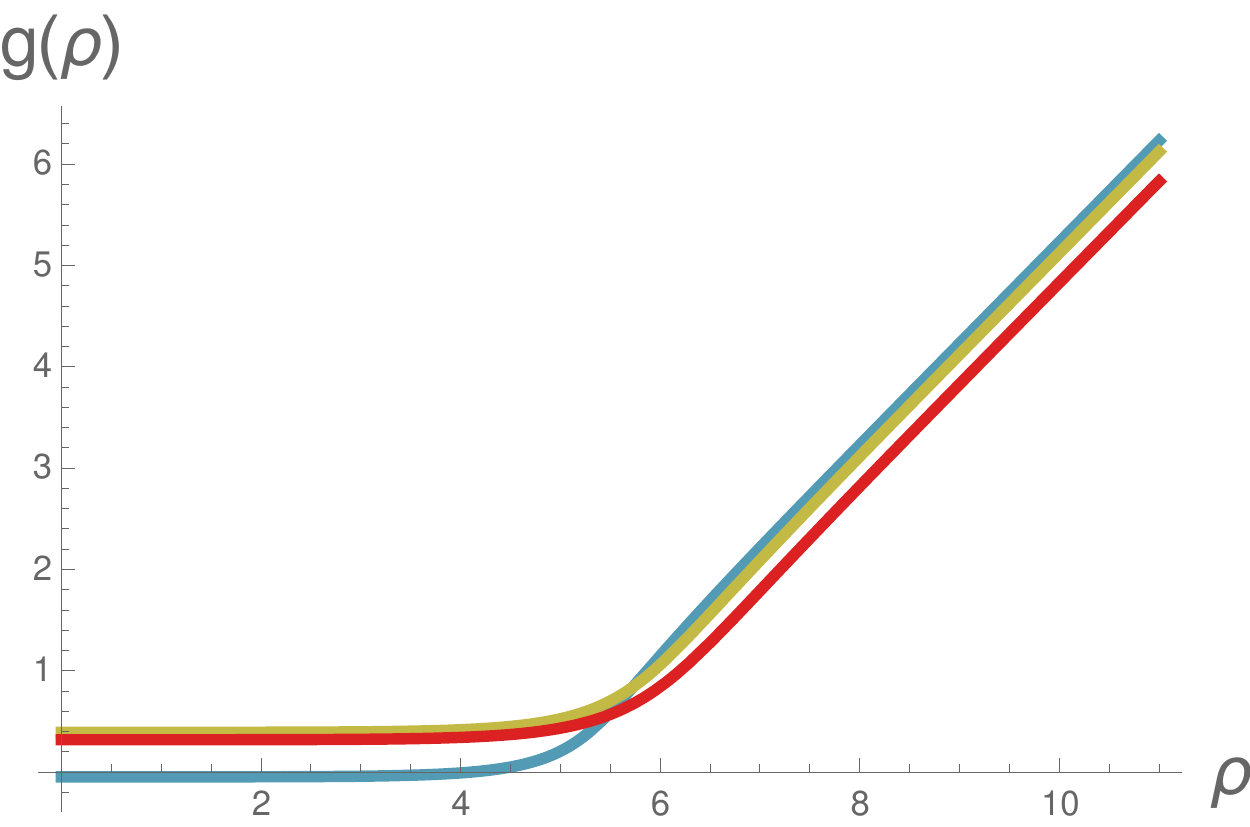}
	\includegraphics[width=0.24\linewidth]{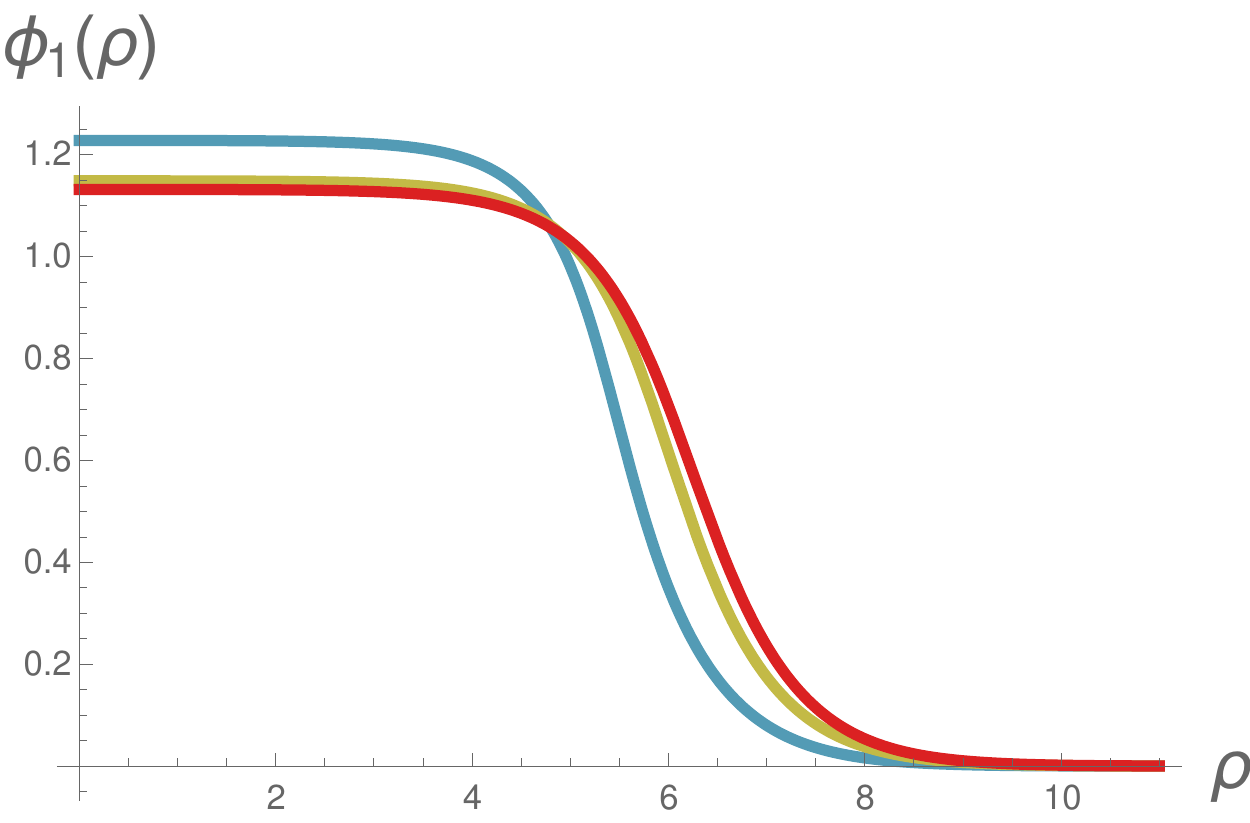}
	\includegraphics[width=0.24\linewidth]{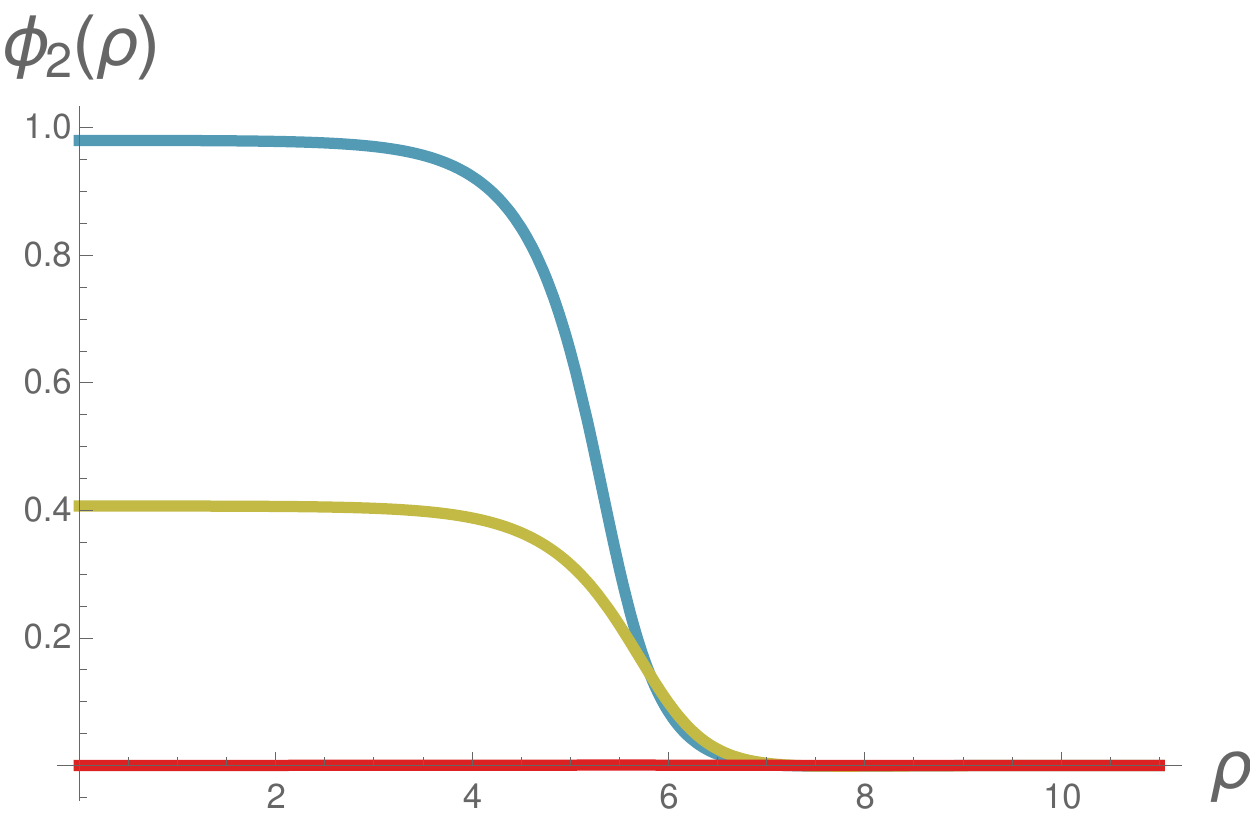}
	\includegraphics[width=0.24\linewidth]{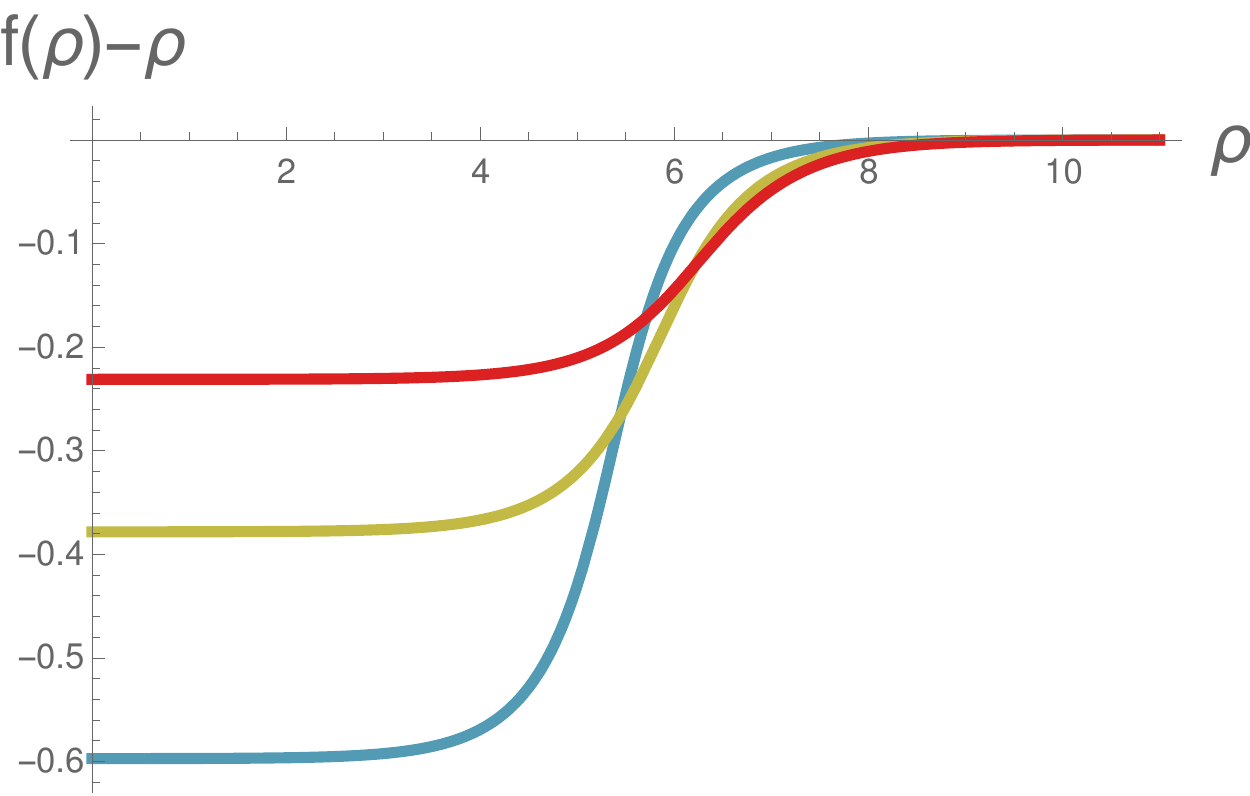}
	
	\caption{$AdS_5\rightarrow AdS_3\times T^2$ flow solutions for $(a_1,a_2)\in\lbrace (1,2), (3,4), (3,3)\rbrace$ in (blue,yellow,red).
		\label{fig:N4flow-sol}}
\end{figure}

We now turn to RG flow solutions, starting with zero temperature. The $AdS_5\rightarrow AdS_3\times T^2$ flow solutions can be constructed numerically. We start from the BPS equations (3.25) in \cite{Benini:2013cda}, for which the radial coordinate is redefined compared to (\ref{eq:ds2-flow-gen}). 
The metric for the flows is written as
\begin{align}\label{eq:ds2-AdS3T2-flow}
	ds^2&=\frac{d\rho^2}{D^2} + e^{2f}ds^2_{\RR^{1,1}} + e^{2g}ds^2_\Sigma~,
\end{align}
where 
\begin{align}
	D&=X^1+\frac{3}{2}a_1X_1~, &
	X^1&=e^{-\frac{\phi_1}{\sqrt{6}}-\frac{\phi_2}{\sqrt{2}}}~, &
	X^2&=e^{-\frac{\phi_1}{\sqrt{6}}+\frac{\phi_2}{\sqrt{2}}}~,
\end{align}
with $X_I=\frac{1}{3}(X^I)^{-1}$ and $X^1X^2X^3=1$. The BPS equations then read
\begin{align}\label{eq:N4SYMBPS}
	D\frac{dg}{d\rho}&=\frac{1}{3}(X^1+X^2+X^3)-e^{-2g}a_I X_I~,
	\nonumber\\
	\frac{D}{\sqrt{6}}\frac{d\phi_1}{d\rho}&=\frac{1}{3}(X^1+X^2-2X^3)+\frac{1}{2}e^{-2g}(a_1X_1+a_2X_2-2a_3 X_3)~,
	\nonumber\\
	\frac{D}{\sqrt{2}}\frac{d\phi_2}{d\rho}&=X^1-X^2+\frac{3}{2}e^{-2g}(a_1X_1-a_2X_2)~.
\end{align}
The remaining function $f$ is determined by
\begin{align}\label{eq:N4-f}
	f&=\rho-\frac{1}{2\sqrt{6}}\phi_1-\frac{1}{2\sqrt{2}}\phi_2~.
\end{align}
Sample solutions for torus compactifications with $a_1+a_2+a_3=0$ are shown in fig.~\ref{fig:N4flow-sol}.
For $a_1=a_2$ the BPS equations can be solved with $\phi_2=0$.
The function $D$ interpolates between $1$ for $\rho\rightarrow \infty$ and constants greater than one for $\rho\rightarrow -\infty$.
In the UV, at $\rho\rightarrow\infty$, $g(\rho)$ and $f(\rho)$ are both linear, leading to exponential warp factors for $\RR^{1,1}$ and $T^2$. The UV asymptotics is $AdS_5$ with $\RR^{1,1}\times T^2$ boundary.
In the IR, at $\rho\rightarrow -\infty$, $g(\rho)$ approaches a constant and decouples from the $AdS$ scaling, while $f$ remains linear, resulting in $AdS_3\times T^2$.
The transition region is around $\rho\approx 5$ in fig.~\ref{fig:N4flow-sol}.

\begin{figure}
	\includegraphics[width=0.28\linewidth]{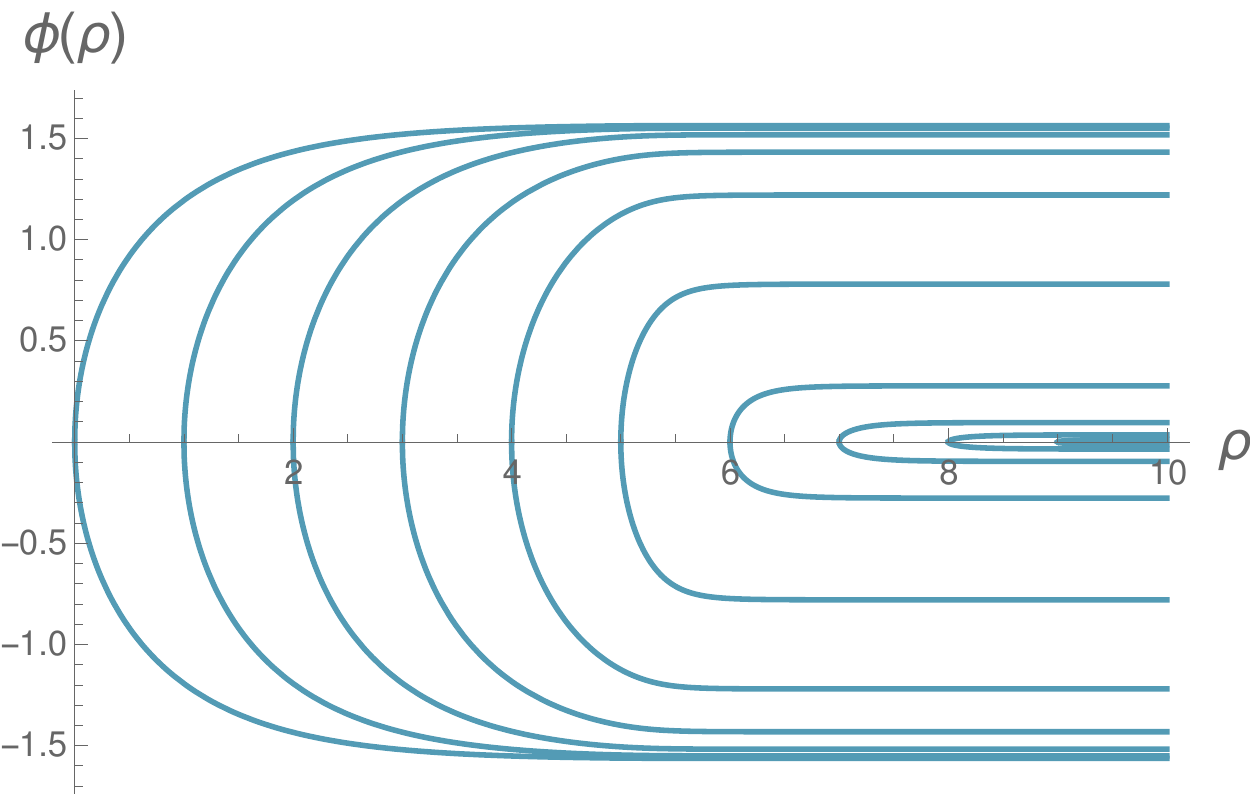}
	\qquad
	\includegraphics[width=0.28\linewidth]{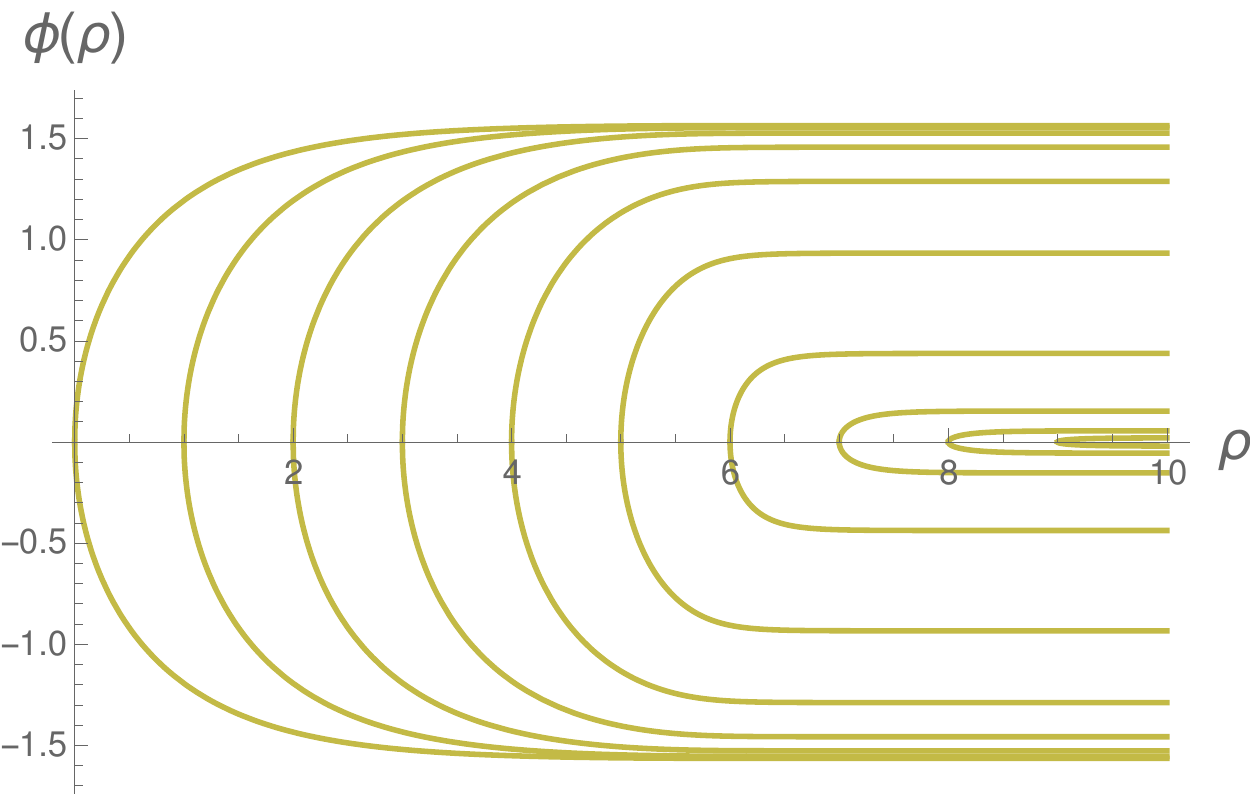}
	\qquad
	\includegraphics[width=0.28\linewidth]{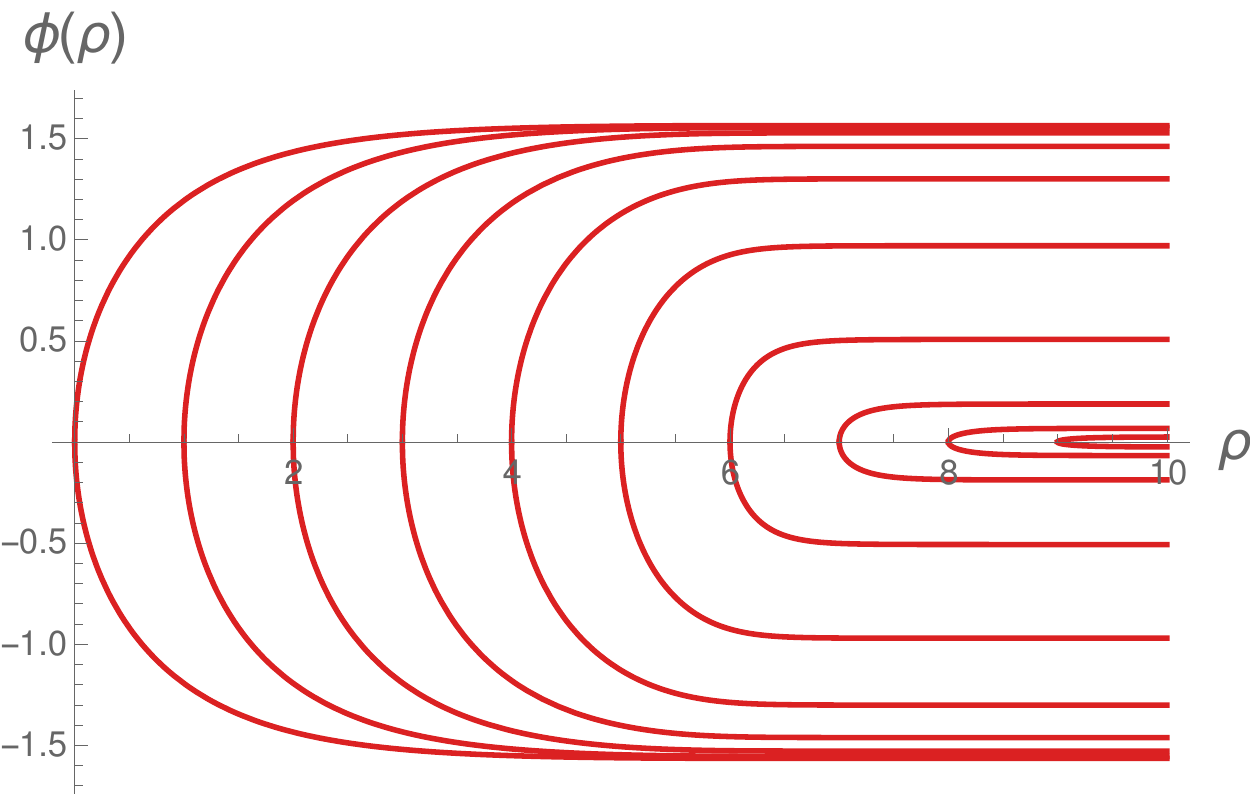}
	\caption{Extremal surfaces in $AdS_5\rightarrow AdS_3\times T^2$ flow solutions of fig.~\ref{fig:N4flow-sol}, in corresponding colors (from left to right for $(a_1,a_2)\in\lbrace (1,2), (3,4), (3,3)\rbrace$).\label{fig:N4flow-surf}}
\end{figure}

In the flow solutions decomposing the torus now leads to geometric EE's.
To match the torus coordinates to those used for the IR fixed point metric (\ref{eq:ds2-AdS3T2}), we choose in (\ref{eq:ds2-AdS3T2-flow})
\begin{align}
	ds^2_\Sigma&=e^{2f_0-2g_0}(d\phi^2+d\chi^2)~,
\end{align}
with the same periodicities for $\phi$ and $\chi$ as in (\ref{eq:ds2-AdS3T2}).
The area of a surface extending along a constant time slice in $AdS_3$, wrapping $S^1_\chi$, and splitting $S^1_\phi$, so that it can be parametrized by $\phi(\rho)$, reads
\begin{align}
	A&=V_{\RR}V_{S^1_\chi}\int d\rho\, e^{f+g+f_0-g_0}\sqrt{\frac{1}{D^{2}}+e^{2g+2f_0-2g_0} \, {\phi^\prime}^2}~.
\end{align}
The equation of motion for $\phi$ leads to 
\begin{align}
	\phi^\prime&=\pm \frac{1}{e^{g+f_0-g_0} D \sqrt{c^2 e^{2f+4g}-1}}~,
\end{align}
with a constant $c^2=e^{-2f-4g}\vert_{\rho=\rho_\star}$ for a surface capping off at $\rho=\rho_\star$.
Integrating for $\phi$ leads to 
\begin{align}
	\phi(\rho)&=\pm \int_{\rho_\star}^\rho \frac{d\hat\rho}{e^{g(\hat\rho)+f_0-g_0}D(\hat\rho)\sqrt{e^{2f(\hat\rho)+4g(\hat\rho)-2f(\rho_\star)-4g(\rho_\star)}-1}}~.
\end{align}
Example surfaces for the flow solutions of fig.~\ref{fig:N4flow-sol} are shown in fig.~\ref{fig:N4flow-surf}.
When the cap-off point $\rho_\star$ is deep in the IR, $\rho_\star\rightarrow -\infty$, the asymptotic value $\phi(\infty)$ becomes independent of $\rho_\star$ and approaches $\pm \pi/2$. This is in line with the observation that, at the IR fixed points, island surfaces at zero temperature only exist for $\phi_0$ given in (\ref{eq:phi0-crit}).
As $\rho_\star$ increases and moves into the transition region towards the UV, the asymptotic value $\phi(\infty)$ decreases. The details depend on the parameters $(a_1,a_2)$, but all solutions support island minimal surfaces also with $\phi_0<\phi_{0,{\rm crit}}$; these surfaces just do not reach into the IR region and can not be seen in the IR fixed point solution.\footnote{Surfaces with constant $\phi=\pm \phi_0$ exist for generic $\phi_0$, so the geometric EE in the UV theory can be computed for arbitrary decompositions of $S_\phi^1$.}
The areas of the surfaces in fig.~\ref{fig:N4flow-surf} depend on $\rho_\star$ already at zero temperature, since conformal invariance is broken by the compactification.
For surfaces reaching deep into the IR the area becomes independent of $\rho_\star$, but for generic $\rho_\star$ shallower surfaces have smaller area.

We now turn to finite temperature. In general, introducing finite temperature in the flow solutions, where the radial coordinate already plays a non-trivial role, is less straightforward than in the IR fixed point solutions.
One has to solve the equations of motion (rather than BPS equations) with appropriate IR boundary conditions.
However, the regime of interest here are temperatures that are small compared to the compactification scale associated with $\Sigma$.
The connection to the UV geometry primarily serves as a form of regulator, and justifies associating EE's with decompositions of the internal space in the IR geometry. In that regime we expect that the solution first flows along the BPS flow from $AdS_5$ in the UV to an intermediate regime where it is well approximated by $AdS_3\times\Sigma$, and then turns into the solution with $AdS_3$ replaced by a black hole in the deep IR.

In the finite temperature flow solutions we expect that island surfaces can be realized for arbitrary $\phi_0$:
Surfaces with $\phi_0\leq \phi_{0,{\rm crit}}$, which cap off in the UV region, are unaffected by the small temperature and take a similar form as at zero temperature (fig.~\ref{fig:N4flow-surf}). Surfaces capping off deep in the IR, on the other hand, take the same form in the IR region as in the IR fixed point solution with finite temperature, where they can be realized for $\phi_0\geq \phi_{0,{\rm crit}}$ (fig.~\ref{fig:surfs-T}). From the IR region they then stretch straight through the UV region.
There is thus no restriction on $\phi_0$ in the finite temperature flow solutions, though not all surfaces reach into the IR.
For surfaces which do reach into the IR, the area difference between island and HM surfaces is dominated by the difference in the IR region, since both stretch approximately straight through the UV region.

The RG flow perspective explains the `tiny island' regime of sec.~\ref{sec:gen}: If one starts in the UV with a $\phi_0$ which is smaller than the critical value (\ref{eq:phi0-crit}), an island surface exists. But upon zooming in on the IR region, e.g.\ by moving a cut-off towards the IR, the island surface recedes towards the boundary of the remaining space and since it does not reach all the way into the IR region it ultimately disappears. In that sense, it leads to an island surface `at the conformal boundary' of the IR geometry.
This is similar to the discussion in \cite{Geng:2020fxl}.

\subsection{\texorpdfstring{Flows to AdS$_5$: 6d $\mathcal N=(2,0)$ on $T^2$}{6d N=(2,0) on T**2}}

Compactifications of M5-branes on surfaces with constant Gaussian curvature $\kappa=\pm 1,0$ and genus $\mathfrak{g}$ were discussed in \cite{Bah:2012dg}, extending the solutions of \cite{Maldacena:2000mw} to incorporate spheres and tori.
The case of interest here is $\mathfrak{g}=1$, discussed in appendix C of \cite{Bah:2012dg}. 
The solutions are obtained in a truncation of maximal 7d gauged supergravity to the metric, two Abelian gauge fields $A_\mu^{(i)}$, two scalars $\lambda_i$ and a 3-form potential which vanishes for the solutions considered here.
The 7d solutions can be uplifted to M-theory, where they describe compactifications of the 6d $\mathcal N=(2,0)$ theories.

The solutions are parametrized by an integer $z\in\ZZ$. The IR fixed point solutions are given by
\begin{align}\label{eq:ds2-M5IR}
	ds^2_7&=e^{2f_0}ds^2_{AdS_5}+e^{2 g_0}(dx_1^2+dx_2^2)~,
	&
	e^{f_0}&=\frac{3^{2/5}}{2^{6/5}}~, & e^{2g_0}&=\frac{3^{3/10}}{2^{2/5}}\frac{|z|}{8}~,
\end{align}
with
\begin{align}
	F^{(1)}_{x_1x_2}&=+\frac{z}{8}~, & \lambda_{1}&=\frac{1}{10}\ln\left(\frac{33}{4}- \frac{19z}{4|z|}\sqrt{3}\right)~,
	\nonumber\\
	 F^{(2)}_{x_1x_2}&=-\frac{z}{8}~, & \lambda_{2}&=\frac{1}{10}\ln\left(\frac{33}{4}+ \frac{19z}{4|z|}\sqrt{3}\right)~.
\end{align}
Flipping the sign of $z$ exchanges $F^{(1)}$ and $F^{(2)}$ as well as $\lambda_1$ and $\lambda_2$.
These solutions preserve 4d $\mathcal N=1$ supersymmetry.
They can be brought into the form of the general metric in (\ref{eq:ds2-gen}) by a rescaling of the torus coordinates.
The EE and Page curve discussion depends again on the volume of the torus, set by $z$, and how it is split.
To match the metric convention to (\ref{eq:ds2-gen}) in the IR we set $(x_1,x_2)=e^{f_0-g_0}(\phi,\chi)$ with the identification $\phi\sim\phi+L$. This fixes the period of $\chi$ in terms of $z$.
Since $d=4$, the Page value $\phi_{0,P}$ in (\ref{eq:phi-P}) differs from the critical value $\phi_{0,{\rm crit}}$ in (\ref{eq:phi0-crit}).

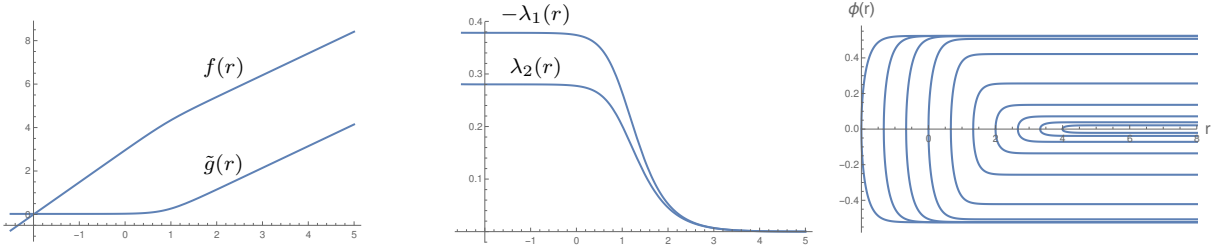
\begin{figure}
	\begin{tikzpicture}
		\node at (0,0) {\includegraphics[width=0.29\linewidth]{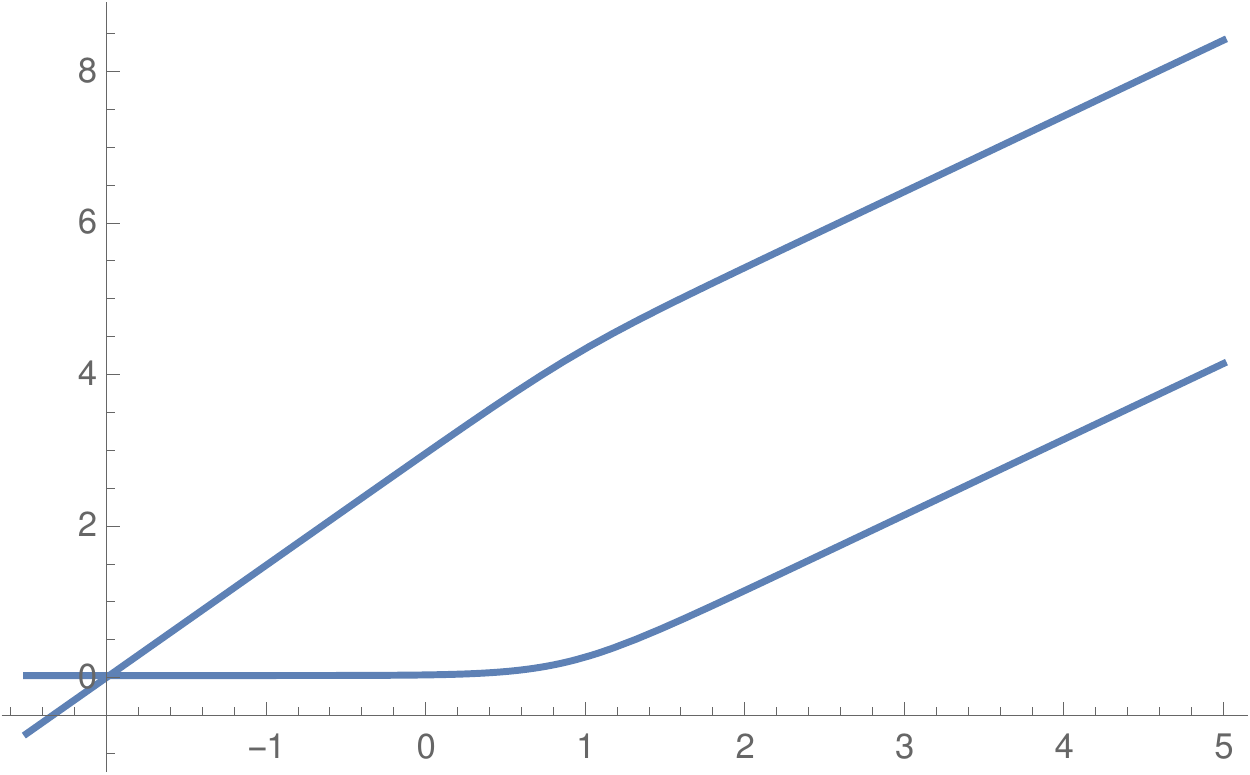}};
		\node at (6,0) {\includegraphics[width=0.29\linewidth]{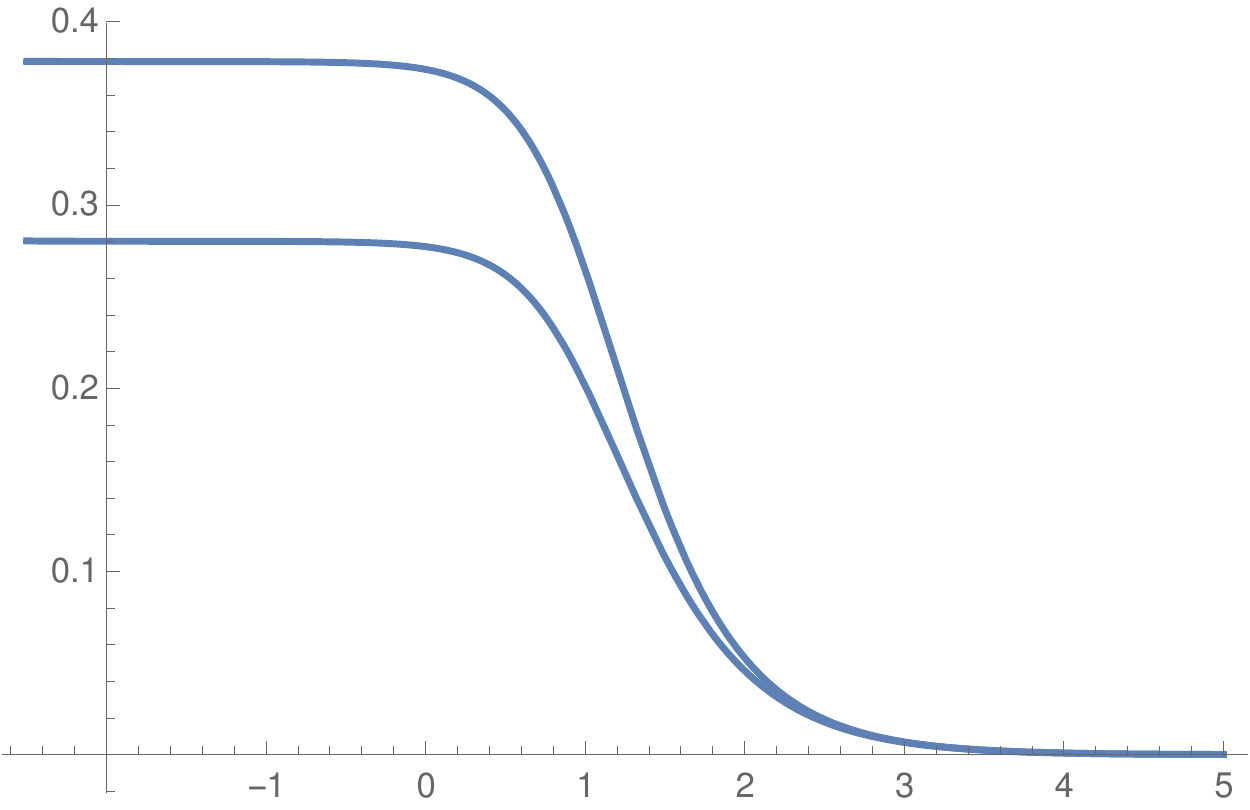}};
		\node at (0.55,0.85) {\scriptsize $f(r)$};
		\node at (0.55,-0.45) {\scriptsize $\tilde g(r)$};
		\node at (4.7,1.55) {\scriptsize $-\lambda_1(r)$};
		\node at (4.7,0.85) {\scriptsize $\lambda_2(r)$};
		
		\node at (11.2,0.2) {\includegraphics[width=0.3\linewidth]{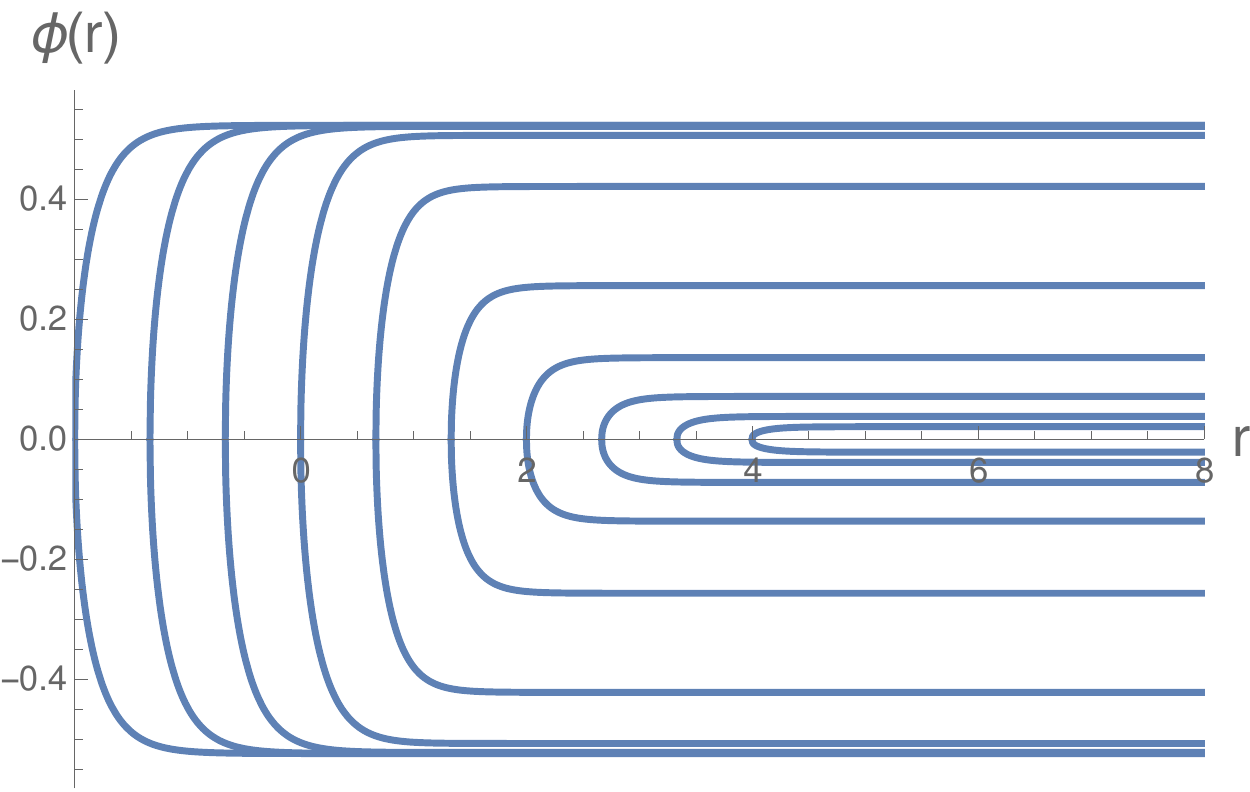}};
	\end{tikzpicture}
	
	\caption{$AdS_7\rightarrow AdS_5\times T^2$ flow solutions. The warp factors are both linear in the UV, corresponding to $AdS_7$ asymptotics, while $\tilde g$ becomes constant in the IR and decouples, leading to $AdS_5\times T^2$. The scalars transition from zero in the UV to non-zero values in the IR. Minimal surfaces are shown on the right. \label{fig:M5flow-sol}}
\end{figure}

The ansatz for flow solutions is that $\lambda_i$ are functions of a radial coordinate $r$, while $F_{x_1,x_2}^{(i)}$ are identical to the IR solution, and the metric takes the form
\begin{align}\label{eq:ds2-M5flow}
	ds^2&=e^{2f}ds^2_{\RR^{1,3}}+e^{2h}dr^2+e^{2\hat g}(dx_1^2+dx_2^2)~,
\end{align}
where $\hat g$, $f$ and $h$ depend only on $r$ for the torus.
The BPS equations were given in (A.4) -- (A.13) of \cite{Bah:2012dg}. 
The non-trivial equations for a torus compactification read (with $m=2$)
\begin{align}
	3\lambda_1'+2\lambda_2'-2 e^{h+2\lambda_1}+2e^{h-4\lambda_1-4\lambda_2}-e^{h-2\hat g-2\lambda_1}F_{x_1x_2}^{(1)}&=0~,
	\nonumber\\
	2\lambda_1'+3\lambda_2'-2 e^{h+2\lambda_2}+2e^{h-4\lambda_1-4\lambda_2}-e^{h-2\hat g-2\lambda_2}F_{x_1x_2}^{(2)}&=0~,
	\nonumber\\
	f'+\lambda_1'+\lambda_2'+e^{h-4\lambda_1-4\lambda_2}&=0~,\label{eq:M5-BPS-3}
	\nonumber\\
	\hat g'-4\lambda_1'-4\lambda_2'+2 e^{h+2\lambda_1}+2e^{h+2\lambda_2}-3e^{h-4\lambda_1-4\lambda_2}&=0~.
\end{align}
The dependence on the fluxes can be eliminated by defining $\hat g=\tilde g+\frac{1}{2}\ln |F_{x_1x_2}^{(1)}| $.
$h$ is a gauge degree of freedom and we choose it such that $e^h dr= - dr$.
When written in terms of $\tilde g$ only a sign variable remains in the BPS equations, encoding the sign of $z$; $z$ itself only appears in the transformation back to the form (\ref{eq:ds2-M5flow}). 
The sign choice only exchanges the two scalars and fluxes.

The numerically obtained solution is shown in fig.~\ref{fig:M5flow-sol}.
Qualitatively the flows are very similar to those for $\mathcal N=4$ SYM in fig.~\ref{fig:N4flow-sol}. In the UV region, at $r\rightarrow\infty$, $f$ and $\tilde g$ both become linear, so that the geometry becomes asymptotically $AdS_7$. In the IR region, at $r\rightarrow -\infty$, $\tilde g$ becomes constant, while $f$ remains linear (with different slope), so that the IR geometry (\ref{eq:ds2-M5IR}) is approached. The scalars turn on in the transition region around $r\approx 1$.

To match the metric convention to (\ref{eq:ds2-gen}) in the IR we again set $(x_1,x_2)= e^{f_0-g_0}(\phi,\chi)$ with $\phi\sim \phi+L$. The area of minimal surfaces splitting the $S^1_\phi$ is then given by
\begin{align}
	A&= V_{\RR^3}V_{S^1_\chi}\int dr e^{3f+\tilde g-\tilde g_0+f_0}\sqrt{1+e^{2\tilde g-2\tilde g_0+2f_0}{\phi'}^2}~.
\end{align}
The only dependence on $z$ is in the length of $S^1_\chi$. The surfaces take a qualitatively similar form to those for the $\mathcal N=4$ SYM flows, as can be seen in fig.~\ref{fig:M5flow-sol}. 
The critical value of $\phi_0$ is $\pi/6$, as predicted by (\ref{eq:phi0-crit}) with $d=4$; the minimal surfaces in fig.~\ref{fig:M5flow-sol} reach to this value when they cap off deep in the IR region.

\subsection{\texorpdfstring{Flows to AdS$_2$ and AdS$_4$}{Flows to AdS2 and AdS4}}\label{sec:ex-contd}

In the previous two sections we discussed RG flows to $AdS_d\times T^2$ with $d=2,4$. In this section we briefly discuss $d=1$ and $d=3$. 
We start with flows to $AdS_2\times T^2$. 
IR fixed point solutions describing wrapped M2 branes were constructed in \cite{Gauntlett:2001qs}.
For $d=1$ extremizing (\ref{eq:area-gen}) at zero temperature leads to linear $\phi(r)$. The condition that a surface should turn around smoothly can not be satisfied. Unlike in $d>1$, only HM surfaces exist at zero temperature.
For $d=1$ and finite temperature an island surface can cap off at $r=r_h$. Extremizing (\ref{eq:area-gen}) leads to
\begin{align}
 \phi(r)&=c\tanh^{-1}\sqrt{b(r)}~.
\end{align}
Since $\phi(r)$ diverges for $r\rightarrow\infty$ the surfaces are self-intersecting unless one introduces a cut-off in $AdS_2$. 
Flow solutions from $AdS_4$ in the UV to $AdS_2\times \Sigma$ in the IR, corresponding to topologically twisted compactifications of ABJM theory on $\Sigma$ or magnetically charged black holes in 4d gauged supergravity with $\Sigma$ horizon, were discussed in \cite{Donos:2011pn} and \cite{Benini:2015eyy}.
Flows to $AdS_2$ in the IR can also be obtained from M5-branes wrapped on two Riemann surfaces, $\Sigma_{\mathfrak{g}_1}\times \Sigma_{\mathfrak{g}_2}$. Such solutions were discussed in \cite{Gauntlett:2001jj} and \cite{Benini:2013cda}.

The perhaps most interesting case is 4d gravity, i.e.\ compactifications leading to $AdS_4\times T^2$.
Solutions for 5d SCFTs compactified on a Riemann surface $\Sigma$ can be constructed in 6d F(4) supergravity \cite{Nunez:2001pt,Naka:2002jz}. They can be uplifted to Type IIA to describe compactifications of the 5d $USp(N)$ theories dual to the Brandhuber/Oz solution \cite{Brandhuber:1999np}. They can also be uplifted to Type IIB, where they describe compactifications of 5-brane web SCFTs dual to the solutions in \cite{DHoker:2016ujz,DHoker:2017mds,DHoker:2017zwj}, using the uplifts \cite{Hong:2018amk,Malek:2018zcz}. This was carried out in \cite{Legramandi:2021aqv}.
However, the solutions in pure 6d F(4) supergravity only accommodate hyperbolic surfaces with genus greater than one.
The twisted compactifications of $\mathcal N=4$ SYM or the 6d $\mathcal N=(2,0)$ theories on $T^2$ use at least two fluxes which can be balanced against each other. The (unique) 5d superconformal algebra $F(4)$, on the other hand, only has an $SU(2)$ R-symmetry and does not offer this option.
Supersymmetric twisted compactifications on tori can be realized by using additional flavor symmetries. 
For the 5d $USp(N)$ theories, which have an $SU(2)_M$ flavor symmetry, holographic duals for torus compactifications were constructed directly in Type IIA in \cite{Bah:2018lyv}. 
The metric of the solutions takes the form
\begin{align}
	ds^2_{10}&=\frac{H^{-1/2}}{\sqrt{y} F_0}\left[ds^2_{AdS_4}+e^{2\nu}ds^2_{\Sigma_{\mathfrak{g}}}+\frac{1}{4}H ds^2_{M_4}\right]~,
	&
	(e^{2\nu})_{\mathfrak{g}=1}&=\frac{|z|}{\sqrt{2}}~,
\end{align}
where $z$ is a parameter specifying the $U(1)_M$ background gauge field, ${M^4}$ is a 4d internal space, and $H$ and $y$ are functions on $M^4$.
The 10d metric is of the general form (\ref{eq:ds2-gen}). 
The solutions describe the IR fixed points, but their existence suggests that flow solutions exist as well.
This provides a string theory uplift of the discussion in sec.~\ref{sec:gen} for $d=3$ corresponding to $AdS_4$.
It would be interesting to study similar compactifications for the landscape of $AdS_6$ solutions in Type IIB.

\section{From geometric to internal entropy on \texorpdfstring{$S^2$}{S**2}}\label{sec:S2}

The topologically twisted compactifications can be used to study the transition from geometric EE's to EE's associated with splits in the internal space more generally. In the following we briefly discuss compactifications on $S^2$ as a model for more general $S^n$ internal spaces in AdS/CFT. 

As shown in \cite{Graham:2014iya}, extremal surfaces which wrap an asymptotically-AdS space and split the internal space necessarily end on an extremal sub-surface in the internal space when reaching the conformal boundary of AdS.
Minimal surfaces splitting an $S^2$ therefore end on an equatorial $S^1$ at the boundary of AdS. This is the only way to split the $S^2$ in the IR geometry.
This result can be motivated as follows. We start with a generic IR fixed point geometry for an $S^2$ compactification
\begin{align}\label{eq:ds2-S2}
	&&
	ds^2_{d+3}&=e^{2f_0}ds^2_{AdS_{d+1}} + e^{2g_0}ds^2_{S^2}~, & ds^2_{S^2}&=d\theta^2+\sin^2\!\theta\,ds^2_{S^1}~.
\end{align}
The $AdS_{d+1}$ factor can be replaced e.g.\ by the black hole metric in (\ref{eq:ds2-bh-d}) without changing the argument.
The area of surfaces parametrized by $\theta(r)$ is
\begin{align}\label{eq:A-S2}
	A&=C\int_{r_\star}^\infty dr\,e^{(d-1)r}\sin\theta\sqrt{\frac{1}{b(r)}+e^{2g_0-2f_0}{\theta^\prime}^2}~,
	&C&=V_{S^1}V_{\RR^{d-1}}e^{df_0+g_0}~.
\end{align}
The UV asymptotics of surfaces approaching $\theta=\frac{\pi}{2}$ for $r\rightarrow \infty$ depends on the ratio of the $S^2$ and $AdS_3$ radii, $e^{2g_0-2f_0}$, but is independent of IR features of the background like the temperature: asymptotically, for large AdS$_{d+1}$ radial coordinate $r$,
\begin{align}\label{eq:Delta-S2}
	\theta(r)&\sim \frac{\pi}{2}+\theta_\pm e^{-\Delta_\pm r}~, & \Delta_\pm&=\frac{1}{2}\left(d-1\pm\sqrt{(d-1)^2-4e^{2f_0-2g_0}}\right)~.
\end{align}
Depending on the (positive) value of $e^{2f_0-2g_0}$, the scaling dimensions $\Delta_\pm$ can either be real and correspond to relevant deformations in the language of AdS$_d$/CFT$_{d-1}$ on the surface, or they can be complex.
The leading behavior of $\theta$ is unchanged either way; small deformations do not change the value of $\theta$ at $r=\infty$. That this extends to the non-linear level was shown in \cite{Graham:2014iya}.
For relevant deformations the surface approaches the equator exponentially, for complex $\Delta_\pm$ it performs damped oscillations around $\theta=\frac{\pi}{2}$ (this will be discussed from the RG flow perspective shortly).

Solving for the actual minimal surfaces leads to two types of solutions: $\theta(r)=\frac{\pi}{2}$ is a simple solution analogous to the HM surface and corresponds to $\theta_\pm=0$ in (\ref{eq:Delta-S2}). More general solutions start at $\theta=\frac{\pi}{2}$ in the UV and cap off at a finite $r_\star$ where the $S^1$ collapses with $\theta\rightarrow 0$ or $\theta\rightarrow \pi$.
The latter are analogous to the island surfaces on $T^2$, and they have non-trivial $\theta_\pm$ in (\ref{eq:Delta-S2}).
The leading divergences in the area for real scaling dimensions are given by
\begin{align}
	A&\approx C\left[\frac{1}{d-1}\frac{1}{\epsilon^{d-1}}-\frac{e^{2f_0-2g_0}\Delta_-\theta_-^2}{2}\frac{1}{\epsilon^{d-1-2\Delta_-}}+\ldots\right],
\end{align}
with a cut-off $r_\epsilon=-\ln \epsilon$. The leading divergence is $\mathcal O(1/\epsilon^{d-1})$ and cancels between the island and HM surfaces. But there are subleading divergences which do not cancel, and the area of the island surfaces is smaller by an infinite amount than the area of the $t=0$ HM surface.

The features of the $S^2$ compactifications are markedly different from $T^2$: For the $T^2$ compactifications there is no restriction on how to split the $T^2$ in the IR geometry. Though at zero temperature island surfaces also only exist for the critical $\phi_0=\phi_{0,{\rm crit}}$ in (\ref{eq:phi0-crit}), straight surfaces with $\phi(r)=\pm \phi_0$  exist for arbitrary $\phi_0$. 
In the flow geometry one can likewise split the $T^2$ arbitrarily.
Since $\phi$ only appears through derivatives in the area functional for $T^2$ in (\ref{eq:area-gen}), fluctuations away from constant $\phi(r)$ correspond to marginal operators with $\Delta_-=0$, $\Delta_+=d-1$ in the language of AdS$_d$/CFT$_{d-1}$ on the surface. As a result, the area difference between HM and island surfaces can be finite, allowing for Page curves on $T^2$. 
If the $S^2$ compactifications are interpreted as model for information transfer from the degrees of freedom represented by one half of the $S^2$ to the other, the island surfaces dominate from the outset and the entropy curve is flat.

\begin{figure}
	\subfigure[][]{\label{fig:S2-flow-1}
		\begin{tikzpicture}
			\node at (0,0) {\includegraphics[width=0.33\linewidth]{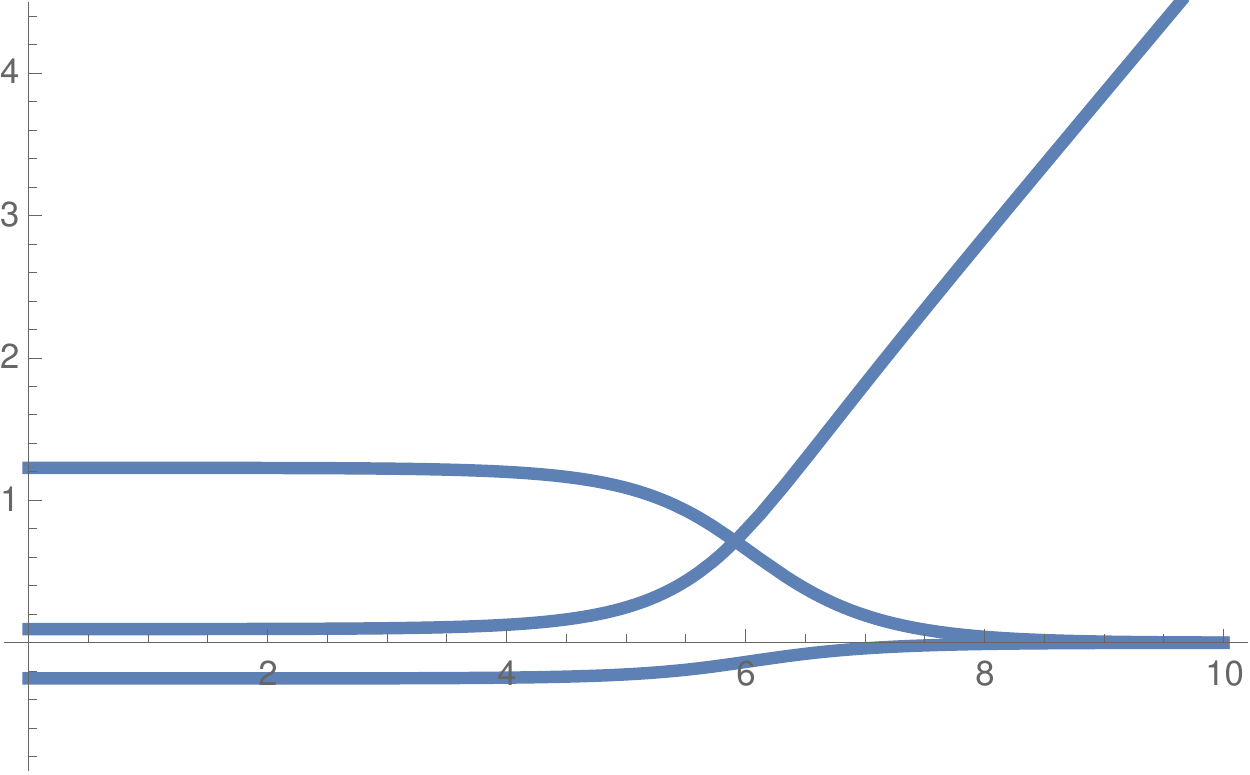}};
			\node at (1.45,1.1) {\footnotesize $g(\rho)$};
			\node at (-1.2,-0.1) {\footnotesize $\phi_1(\rho)$};
			\node at (-1.2,-1.5) {\footnotesize $f(\rho)-\rho$};
		\end{tikzpicture}
	}\hskip 20mm
	\subfigure[][]{\label{fig:S2-1}
		\includegraphics[width=0.33\linewidth]{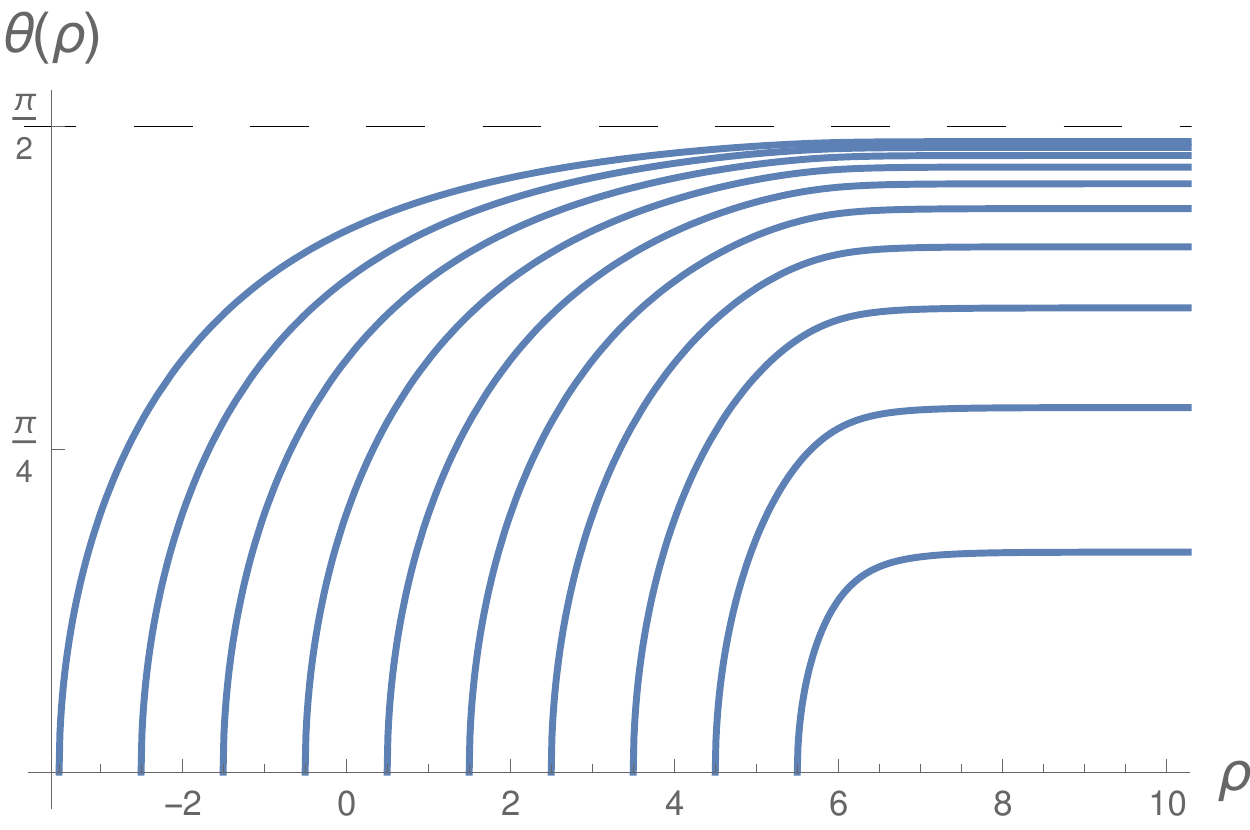}
	}
	\caption{Left: RG flow solution for $\mathcal N=4$ SYM on $S^2$ with $a_1=a_2=2$. Right: Minimal surfaces; surfaces starting close to the equator in the UV reach into the IR. The scaling dimensions in the IR solution are real.\label{fig:N4-S2-real}}
\end{figure}
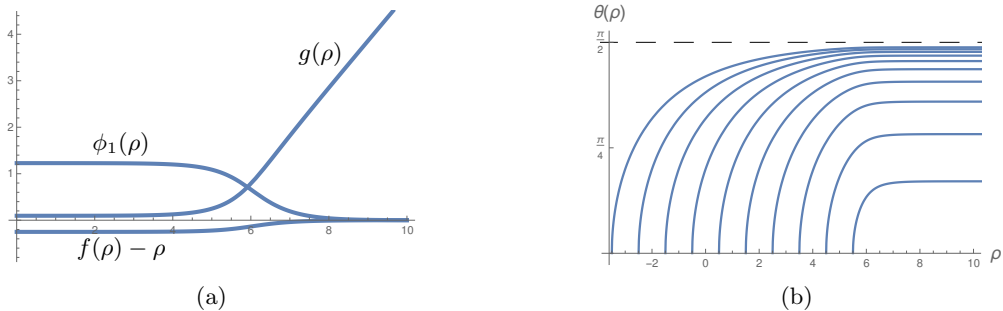

We now discuss the RG flow perspective on the restriction to $\theta\rightarrow \frac{\pi}{2}$ in the UV and on the scaling dimensions. 
As example we choose $\mathcal N=4$ SYM, for which the flows were discussed in \cite{Benini:2013cda}. 
The metric of the IR fixed point solution is given by (\ref{eq:ds2-S2}) with $d=2$; $g_0$, $f_0$ are given in (\ref{eq:ThetaPi}) and the scalars are in (\ref{eq:N4SYM-IR1}).
Regular $AdS_3\times S^2$ solutions exist if the fluxes are such that two of the $a_I$ are positive and the remaining one, which is constrained by 
\begin{align}\label{eq:flux-S2}
	a_1+a_2+a_3=-1~,
\end{align}
is negative. 
We note further that the $a_I$ are half-integer quantized for $S^2$.
The scaling dimensions $\Delta_\pm$ in (\ref{eq:Delta-S2}) are real if at least one of the two positive $a_I$ is greater than $2$, and complex otherwise.

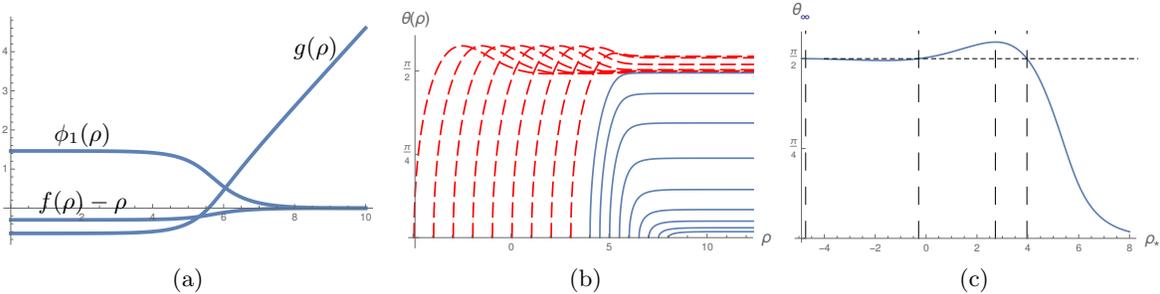
\begin{figure}
	\subfigure[][]{\label{fig:S2-flow-2}
		\begin{tikzpicture}
			\node at (0,0) {\includegraphics[width=0.3\linewidth]{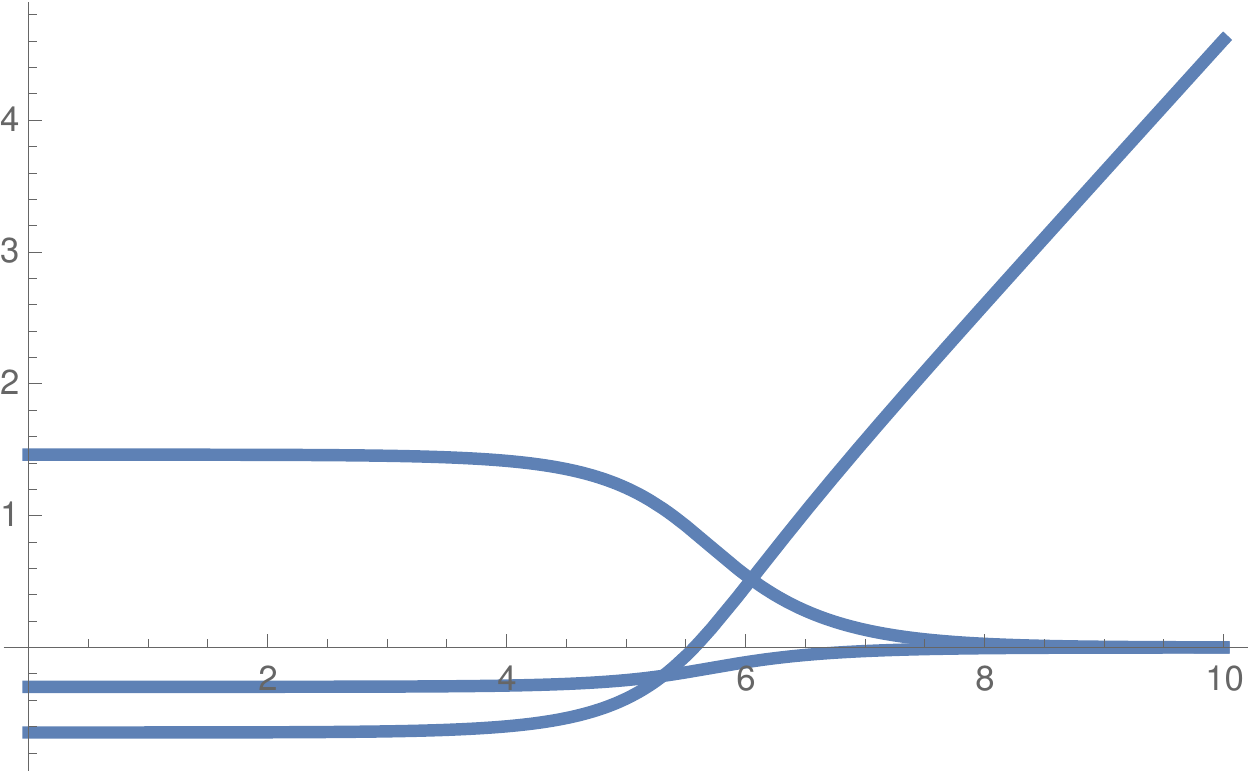}};
			\node at (1.7,1.1) {\footnotesize $g(\rho)$};
			\node at (-1.4,-0.05) {\footnotesize $\phi_1(\rho)$};
			\node at (-1.4,-0.95) {\footnotesize $f(\rho)-\rho$};
		\end{tikzpicture}
	}\hskip 0mm
	\subfigure[][]{\label{fig:S2-2}
		\includegraphics[width=0.3\linewidth]{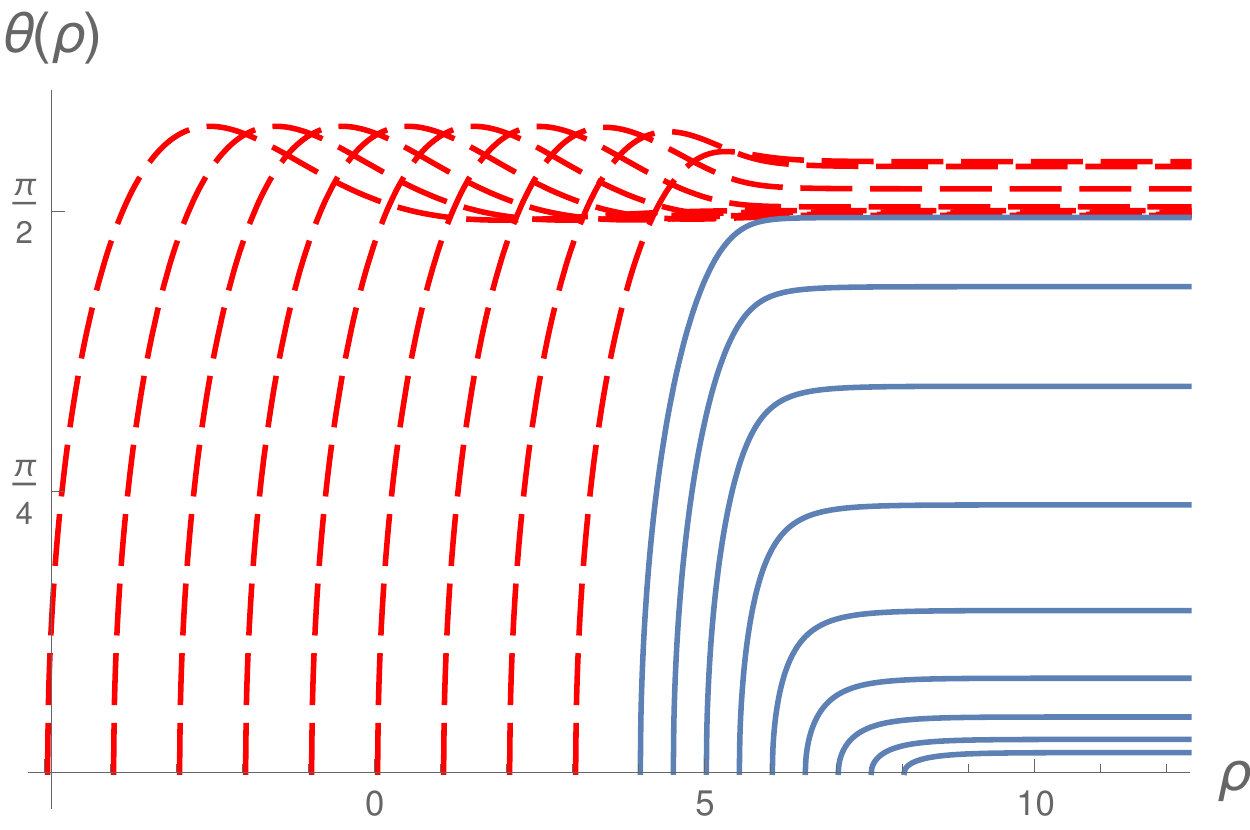}
	}\hskip 0mm
	\subfigure[][]{\label{fig:S2-2-bc}
		\includegraphics[width=0.3\linewidth]{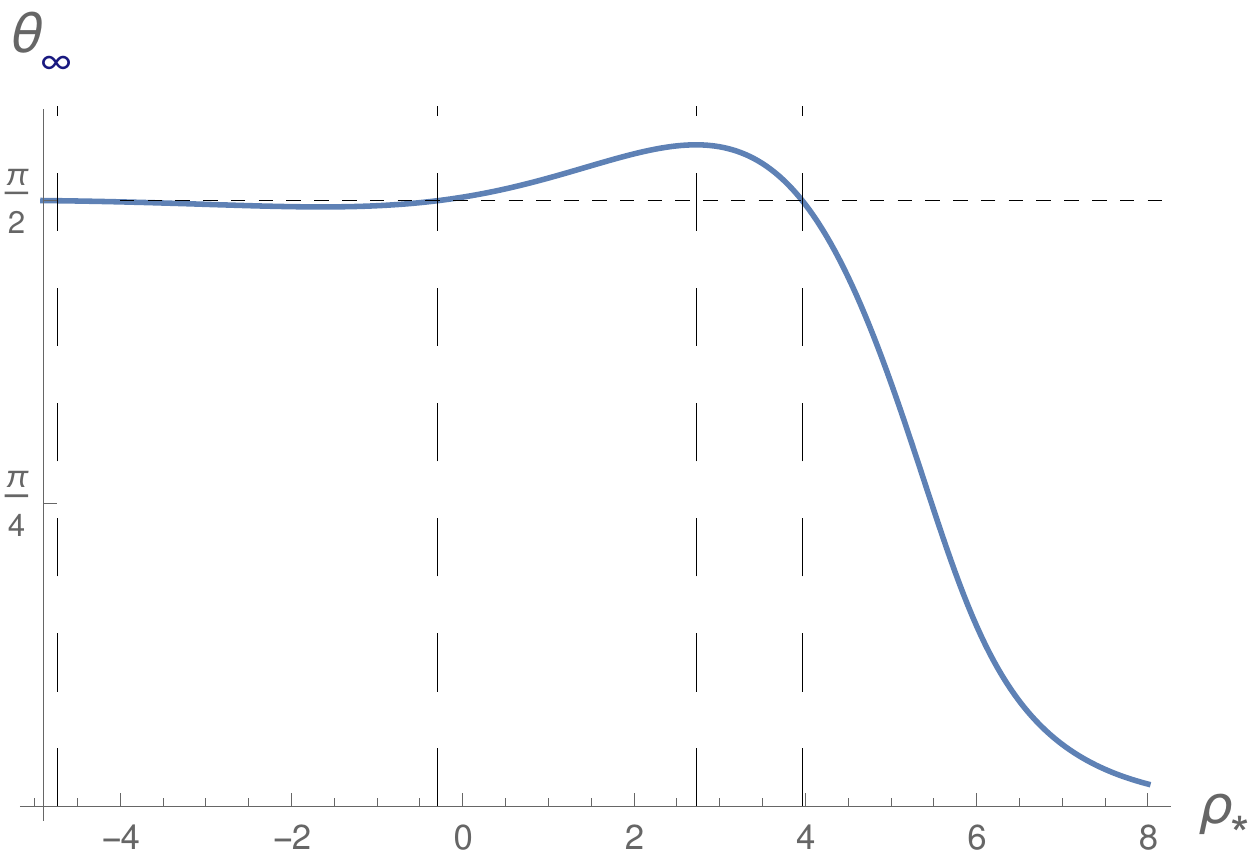}
	}
	\caption{Left: RG flow solution for $\mathcal N=4$ SYM on $S^2$ with $a_1=a_2=\frac{1}{2}$. The IR scaling dimensions are complex. Center: Extremal surfaces. The red dashed curves are extremal surfaces that reach beyond $\frac{\pi}{2}$. The blue surfaces are minimal surfaces. Right: Anchor point $\theta_\infty\equiv \theta(\infty)$ as function of $\rho_\star$.\label{fig:N4-S2-complex}}
\end{figure}

As metric for the flow solutions (at zero temperature) we take, analogously to (\ref{eq:ds2-AdS3T2-flow}),
\begin{align}
	ds^2&=\frac{d\rho^2}{D^2}+e^{2f(\rho)}ds^2_{\RR^{1,1}}+e^{2g(\rho)}ds^2_{S^2}~.
\end{align}
The BPS equations are given by (\ref{eq:N4SYMBPS}), (\ref{eq:N4-f}), with the fluxes now constrained by (\ref{eq:flux-S2}).
For $a_1=a_2$ the equations can again be solved with $\phi_2=0$ and we focus on those cases for the examples.
Explicit solutions obtained numerically are shown in figs.~\ref{fig:S2-flow-1} and \ref{fig:S2-flow-2}.
In these plots the UV/IR limits in the RG flow solutions correspond to large positive/negative $\rho$; the transition region from $AdS_5$ to $AdS_3\times S^2$ is around $\rho\sim 5$.
The area of the surfaces splitting the $S^2$ becomes
\begin{align}
	A&=\Vol_{\RR}V_{S^1}\int d\rho\, e^{f+g}\sin\theta(\rho)\sqrt{\frac{1}{D^{2}}+e^{2g} \, {\theta^\prime}^2}~.
\end{align}
The $\theta(r)=\frac{\pi}{2}$ surface is a solution in the full RG flow background.
More general solutions to the extremality conditions are shown in figs.~\ref{fig:S2-1} and \ref{fig:S2-2}. 
In the UV the $S^2$ is part of the field theory geometry and can be decomposed arbitrarily -- there are no restrictions on the values of $\theta$ at which the minimal surfaces can be anchored at $\rho\rightarrow\infty$. Whether and how the surfaces reach into the IR region depends on the fluxes $a_1$, $a_2$.

For the example in fig.~\ref{fig:N4-S2-real} with $a_1=a_2=2$ the scaling dimensions in the IR solution are real. Surfaces starting at a generic non-equatorial $S^1$ in the UV cap off before reaching the IR region of the geometry and can not be seen in the IR fixed point solution, in line with the discussion above.
These surfaces are similar to the tiny island surfaces on $T^2$.
Only surfaces starting close to the equator reach through the transition region into the IR geometry.
Surfaces capping off in the deep IR limit of the geometry have to start infinitesimally close to the equator in the UV, and lead to the behavior in (\ref{eq:Delta-S2}) from the IR perspective.
The complete set of surfaces comprises those in fig.~\ref{fig:S2-1} and those obtained by the replacement $\theta(\rho)\rightarrow \pi-\theta(\rho)$.

For the example in fig.~\ref{fig:N4-S2-complex} with $a_1=a_2=\frac{1}{2}$, on the other hand, the scaling dimensions in the IR solution are complex.
From the perspective of the IR fixed point solution, the pair of complex scaling dimensions in (\ref{eq:Delta-S2}) leads to surfaces which in the UV limit of the IR solution perform damped oscillations around $\theta=\frac{\pi}{2}$. This behavior manifests itself in the flow solutions as well.
Fig.~\ref{fig:S2-2} shows two kinds of surfaces: 
\begin{itemize}
	\item[\bf --] Surfaces starting at a value $\theta_\infty\equiv\theta(\rho=\infty)<\frac{\pi}{2}$ at the conformal boundary and capping off with $\theta\rightarrow 0$ at some finite $\rho_\star$. These surfaces are qualitatively similar to the surfaces in fig.~\ref{fig:S2-1}, except that they do not reach deep into the IR region even as $\theta_\infty$ approaches $\frac{\pi}{2}$. They instead always cap off in the UV, before reaching a critical value $\rho_{\rm crit}$.
	\item[\bf --] Surfaces anchored at $\theta_\infty>\frac{\pi}{2}$ in the UV, which reach beyond $\rho_{\rm crit}$ into the IR before capping off with $\theta\rightarrow 0$. The anchor point $\theta_\infty$ as function of the cap-off point $\rho_\star$ is shown in fig.~\ref{fig:S2-2-bc}. $\theta_\infty$ initially increases as $\rho_\star$ is decreased below $\rho_{\rm crit}$, up to a value $\rho_{\rm max}$ where $\theta_\infty$ takes a maximum. Upon decreasing $\rho_\star$ further, $\theta_\infty$ approaches $\frac{\pi}{2}$ through damped oscillations.
\end{itemize}
The complete set of surfaces again comprises those in fig.~\ref{fig:S2-2} and those obtained by the replacement $\theta(\rho)\rightarrow \pi-\theta(\rho)$.
The first type of surfaces does not reach into the IR region, in line with the constraint discussed above.
The surfaces of the second class can reach into the IR, and they approach their limiting value $\theta_\infty$ through damped oscillations for $\rho_\star<\rho_{\rm max}$. 
This matches the behavior deduced from the surfaces in the IR fixed point solution with complex scaling dimensions.

For a range of anchor points $\theta_\infty$ around $\frac{\pi}{2}$ there are multiple extremal surfaces ending on the same $S^1$ at the conformal boundary. However, the surfaces reaching further into the IR have larger area than the surfaces capping off further in the UV.
That means only the surfaces shown as solid lines in fig.~\ref{fig:S2-2} are actual minimal surfaces.
The extremal surfaces (in the sense that the first variation vanishes) that reach all the way into the IR are subdominant with respect to surfaces anchored at the same $\theta_\infty$ and capping off further in the UV.
The surfaces that are relevant for computing the EE (at least at leading order) do not reach into the IR.
Similar {\it entanglement shadows} were noted in \cite{Balasubramanian:2014sra,Balasubramanian:2017hgy}.
Whether the non-minimal extremal surfaces reaching into the IR can be related to field theory quantities would be interesting to understand (for non-minimal extremal surfaces in AdS$_3$ an interpretation in terms of {\it entwinement} was proposed in \cite{Balasubramanian:2014sra}).

It would more generally be interesting to understand the qualitatively different behavior of the minimal surfaces for different values of the flux parameters $a_I$ in the twisted compactifications on $S^2$ from the field theory side.	
In AdS/CFT more generally, cases where the scaling dimensions for surfaces splitting spheres in the internal space are complex are not uncommon, and include surfaces splitting the $S^5$ in $AdS_5\times S^5$ in Type IIB \cite{Karch:2014pma} and the surfaces of \cite{Anous:2019rqb}.
On the other hand, the $AdS_3\times S^3\times S^3\times S^1$ solutions \cite{Gukov:2004ym,Eberhardt:2017pty} are an example where the radius of each of the $S^3$'s can take a range of values depending on the choice of brane charges, so that real and complex scaling dimensions for surfaces splitting one of the $S^3$'s can both be realized.
The discussion above suggests that surfaces splitting an internal $S^3$ with real scaling dimensions should have an interpretation as EE between non-geometrically defined subsystems, while the interpretation in the cases with complex scaling dimensions may be more subtle.

Finally, it would be interesting to study compactifications on hyperbolic surfaces, where splits in the IR geometry are again only allowed along extremal sub-surfaces. 
Hyperbolic surfaces can be realized as quotients of the Poincar\'e disc, and the metric and area functional can be obtained locally from (\ref{eq:ds2-S2}) and (\ref{eq:A-S2}) by replacing $\sin\theta \rightarrow \sinh\theta$. This leads to real scaling dimensions corresponding to irrelevant operators.  We leave a more detailed investigation of the transition from geometric to internal EE's along RG flows for the future. The spin-2 spectrum for surfaces which are almost split was discussed recently in \cite{DeLuca:2021ojx}.

\let\oldaddcontentsline\addcontentsline
\renewcommand{\addcontentsline}[3]{}
\begin{acknowledgments}
I am grateful to Andreas Karch, Jim Liu and Leo Pando-Zayas for interesting discussions and correspondence.
This work is supported, in part, by the US Department of Energy under Grant No.~DE-SC0007859 and by the Leinweber Center for Theoretical Physics.
\end{acknowledgments}
\let\addcontentsline\oldaddcontentsline

\bibliography{gravitatingbath}
\end{document}